\documentclass[11pt]{article}
\usepackage[normalem]{ulem}
\usepackage{hyperref}
\usepackage{amsmath}
\usepackage{amsfonts}
\usepackage{braket}
\usepackage{amsthm}
\usepackage{amssymb}
\usepackage{url}
\usepackage{algorithm}
\usepackage{algpseudocode}
\usepackage{mathtools}
\usepackage[title]{appendix}
\usepackage[all]{xy}
\usepackage{geometry}
\usepackage{physics}
\usepackage{cancel}
\usepackage{mdframed}
\usepackage[most]{tcolorbox}
\usepackage{enumitem}
\usepackage{xcolor}
\usepackage{comment}
\usepackage{array}
\usepackage{stackengine,scalerel}
\allowdisplaybreaks

\newtheorem{theorem}{Theorem}[section]

\newtheorem{lemma}[theorem]{Lemma}
\newtheorem{proposition}[theorem]{Proposition}

\newtheorem*{conjecture*}{\textbf{Conjecture}}
\newtheorem{corollary}[theorem]{Corollary}

\newtheorem{claim}{Claim}[theorem]
\newtheorem{fact}{Fact}[theorem]

\newcounter{def}
\theoremstyle{definition}
\newtheorem{definition}{Definition} 

\newtheorem{notation}{\textbf{Notation}}
\newtheorem{remark}[theorem]{Remark}
\newtheorem*{remark*}{\textbf{Remark}}

\newcommand{\CheckCoins}{\mathsf{CheckCoins}}
\newcommand{\SoftDecision}{\mathsf{SoftDecision}}
\newcommand{\SoftDecisionProj}{\mathsf{SoftDecisionProj}}

\newcommand{\ValEst}{\mathsf{ValEst}}
\newcommand{\Prepare}{\mathsf{Prepare}}
\newcommand{\Repair}{\mathsf{Repair}}
\newcommand{\Accept}{\mathsf{Accept}}

\newcommand{\iter}{\mathsf{iter}}

\newcommand{\negl}{\mathsf{negl}}
\newcommand{\bfrho}{\boldsymbol{\rho}}

\newcommand{\eps}{\varepsilon}
\newcommand{\poly}{\text{poly}}

\newcommand{\TD}{\mathsf{TD}}
\newcommand{\secp}{\lambda}
\newcommand{\QPT}{\mathsf{QPT}}

\newcommand{\brho}{\boldsymbol{\rho}}
\newcommand{\bsigma}{\boldsymbol{\sigma}}
\newcommand{\PPT}{\mathsf{PPT}}

\newcommand{\overbar}[1]{\mkern 1.5mu\overline{\mkern-1.5mu#1\mkern-1.5mu}\mkern 1.5mu}
  
\newcommand{\labeleq}{\stepcounter{equation}\tag{\theequation}}

\makeatletter
\newcommand{\customlabel}[2]{%
   \protected@write \@auxout {}{\string \newlabel {#1}{{#2}{\thepage}{#2}{#1}{}} }%
   \hypertarget{#1}{#2}
}
\makeatother

\definecolor{classicrose}{rgb}{0.98, 0.8, 0.91}
\newcommand{\algo}[3]{
    \stepcounter{figure}
    \vspace{0.15cm}
    { \small
    \begin{tcolorbox}[breakable, enhanced, colback=classicrose!20]
    \begin{center}
    {\bf \underline{Algorithm~\customlabel{alg:#2}{\thefigure}: #1}}
    \end{center}
    
    #3
    \end{tcolorbox}
    }
}

\newcommand{\defbox}[4]{
    \stepcounter{def}
    \vspace{0.15cm}
    { \small
    \begin{tcolorbox}[breakable, enhanced]
    \begin{center}
    {\bf \underline{{#1}~\customlabel{defbox:#3}{\thedef}: #2}}
    \end{center}
    
    #4
    \end{tcolorbox}
    }
}

\newcommand{\calA}{\mathcal A}
\newcommand{\calB}{\mathcal B}

\newcommand{\calD}{\mathcal D}
\newcommand{\calE}{\mathcal E}
\newcommand{\calF}{\mathcal F}

\newcommand{\calH}{\mathcal H}
\newcommand{\calI}{\mathcal I}

\newcommand{\calM}{\mathcal M}

\newcommand{\calP}{\mathcal P}

\newcommand{\calZ}{\mathcal Z}

\newcommand{\sfM}{\mathsf M}

\newcommand{\bbE}{\mathbb E}

\newcommand{\bbN}{\mathbb N}

\title{Parallel Repetition for Post-Quantum Arguments}
\author{Andrew Huang \\
MIT CSAIL \\
\texttt{ahuang@mit.edu}
\and
Yael Tauman Kalai \\
MIT CSAIL \\
\texttt{tauman@mit.edu}
}
\date{June 2, 2025}
\begin{document}
\maketitle
\begin{abstract}
In this work, we show that parallel repetition of \emph{public-coin} interactive arguments reduces the soundness error at an exponential rate even in the \emph{post-quantum} setting. Moreover, we generalize this result to hold for threshold verifiers, where the parallel repeated verifier accepts if and only if at least $t$ of the executions are accepted (for some threshold~$t$). Prior to this work, these results were known only when the cheating prover was assumed to be classical.

We also prove a similar result for three-message \emph{private-coin} arguments. Previously, Bostanci, Qian, Spooner, and Yuen (STOC 2024) proved such a parallel repetition result in the more general setting of quantum protocols, where the verifier and communication may be quantum. We consider only protocols where the verifier is classical, but obtain a simplified analysis, and for the more general setting of threshold verifiers.
\end{abstract}
\newpage

\enlargethispage{1cm}
\tableofcontents
\pagenumbering{roman}
\newpage
\pagenumbering{arabic}

\section{Introduction}\label{sec:intro}
One of the most important techniques in the study of interactive proofs, probabilistically checkable proofs, and multi-prover interactive proofs is that of \emph{soundness amplification}, where the goal is to improve soundness (i.e., reduce the probability that a malicious prover makes the verifier accept a false statement). This has also been studied in the context of interactive arguments, which are interactive protocols that are guaranteed to be sound only against computationally bounded cheating provers. We focus on interactive arguments in the \emph{post-quantum} setting, where the verifier (and thus the communication) remains classical but soundness is required to hold even against quantum adversaries of polynomial size. 

One natural way to amplify soundness is via sequential repetition, namely, by repeating the protocol sequentially, where the verifier accepts only if all sequential executions are accepted. Indeed, this transformation is known to reduce the soundness error at an exponential rate, but it has the undesirable property of significantly increasing the round complexity of the original protocol. A preferable way to reduce the soundness error would be via \emph{parallel repetition}, as it preserves the round complexity of the original protocol. However, quite surprisingly, in the context of interactive arguments, parallel repetition does not always reduce the soundness error \cite{BIN97}.

\paragraph{The classical setting.} The question of parallel repetition has been studied quite extensively in the classical setting. It is a well-known fact that parallel repetition decreases the soundness error exponentially for proofs with statistical soundness \cite{BM88, KW00}. However, for proofs with computational soundness, the situation is much more complicated. 

The study of parallel repetition in the computationally sound setting dates back to the work of Yao \cite{Yao82} on hardness amplification of one-way functions, which can be viewed as establishing that parallel repetition reduces the soundness error at an asymptotically optimal rate in every publicly-verifiable two-message argument. Later, Bellare, Impagliazzo, and Naor \cite{BIN97} extended this result by proving that parallel repetition reduces the soundness error exponentially for any three-message interactive argument. Quite surprisingly, in that same paper, Bellare et al.\ also showed that this is not the case for four-message arguments, by constructing a family of four-message interactive arguments (assuming the existence of a non-malleable encryption scheme), where for each $k$, there exists a protocol whose $k$-fold parallel repetition has the same soundness error.\footnote{The communication complexity of a single-fold execution in their family of argument systems grows linearly with $k$. Later work by Pietrzak and Wikstrom \cite{PW07} constructed an interactive argument (from standard computational assumptions) where the $k$-fold parallel repetition does not decrease the error probability below some constant for any polynomial $k$, and where the communication complexity of a single-fold execution does not depend on $k$.} Loosely speaking, their interactive arguments are constructed to allow a cheating prover to coordinate among the different executions. More specifically, their cheating prover cheats by using the verifier's messages from other executions. Importantly, the verifier's messages are private-coin; otherwise they would be useless, as the cheating prover could have simulated them on her own.

Indeed, it was proven that for public-coin protocols,\footnote{These are protocols where all the verifier's messages are computed by simply tossing random coins and sending the outcome.} and more generally for simulatable protocols,\footnote{These are protocols where the verifier's messages are efficiently and computationally simulatable, but the verifier's verdict may still require private randomness.} parallel repetition reduces the soundness error exponentially \cite{PV07,HPPW10,CL10,CP15}. All these works left open the following question: \emph{for which multi-round private-coin protocols does parallel repetition reduce the soundness error, and at what rate?}

Haitner \cite{Hai09} was the first to show a generic transformation that converts any interactive argument into one for which parallel repetition decreases the soundness error at an exponential rate. Moreover, his transformation, known as \emph{random-termination}, is extremely simple and preserves the round complexity of the original protocol. The transformation only slightly modifies the protocol by having the verifier terminate the protocol in each round with some small probability (in which case the verifier automatically accepts). Haitner proved that while this slightly increases the soundness error of the protocol, the soundness error decreases exponentially when repeated in parallel! 

We emphasize that all the works mentioned above were in the classical setting and considered only classical cheating provers. Indeed, the proofs of these parallel repetition theorems proceeded by converting a cheating prover for the parallel-repeated protocol into a cheating prover for a single execution, by rewinding the parallel prover over and over again. We note that such (perfect) rewinding \emph{cannot} be done in the quantum setting due to the no-cloning principle and the destructive nature of quantum measurements. Given the increasing progress towards building large-scale quantum computers, it is of utmost importance to ensure the security of our systems and soundness of our proofs against quantum cheating provers. 

\paragraph{The quantum setting.}
A recent work by Bostanci, Qian, Spooner, and Yuen \cite{BQSY24} considers the \emph{quantum} setting where both the prover and verifier have quantum capabilities. They show that for three-message quantum argument systems, parallel repetition decreases soundness exponentially. They also demonstrate a round compression procedure which converts any quantum argument system with a polynomial number of rounds to a three-message, public-coin quantum argument system with polynomially-related soundness.

Importantly, this round compression procedure is unlikely to have a classical analogue, even for proof systems: while it is known that $\mathsf{QIP}(3) = \mathsf{QIP}$ \cite{KW00}, it is widely believed that $\mathsf{IP}(3) \neq \mathsf{IP}$, as otherwise $\mathsf{AM} = \mathsf{PSPACE}$, which implies the collapse of the polynomial hierarchy.

\subsection{Our Results}
In this work, we consider the \emph{post-quantum} setting, where both the verifier and communication are restricted to being classical, but soundness is required to hold even against quantum cheating provers. 
We establish parallel repetition theorems for two classes of post-quantum interactive arguments. 

First, we show that for public-coin protocols with an arbitrary polynomial number of rounds, parallel repetition decreases the soundness error at an exponential rate. 
\begin{theorem}[Informal]\label{thm:informal1}
    For any public-coin $m$-message protocol with soundness error $s$, its $k$-fold parallel repetition has soundness error $\leq  f(s)^{O(k/m^2)}+\negl(\secp)$, where $f(s) \approx 2^{-(1-s)^2}$.
\end{theorem}
Our result is lossy in two aspects: first, the soundness error goes down exponentially in $k/m^2$ as opposed to $k$. Secondly, the base is $f(s)$ and opposed to $s$. We note that a similar soundness loss was given in the classical setting in \cite{HPPW10} and \cite{CP15}. We generalize Theorem \ref{thm:informal1} by considering the setting of threshold verifiers, who accept if the number of accepting transcripts among all executions exceeds some threshold.  
\begin{theorem}[Informal]\label{thm:informal2}
    For any public-coin $m$-message protocol with soundness error $s$, its $k$-fold parallel repetition with threshold $0 \leq t \leq k$ has soundness error $\leq f(s)^{O(k/m^2)}+\negl(\secp)$, where $f(s) \approx 2^{-\left(t/k-s\right)^2}$.
\end{theorem}

Second, we show a parallel repetition theorem for any three-message (possibly private-coin) protocol.
\begin{theorem}[Informal] \label{thm:informal3}
    For any $3$-message protocol with soundness error $s$, its $k$-fold parallel repetition has soundness error $\leq f(s)^{O(k)}+\negl(\secp)$, where $f(s) \approx 2^{-(1-o_k(1)-s)^2}$.\footnote{It should be noted that \cite{BQSY24} provides a tighter result in the three-message setting. We give a different proof of parallel repetition which extends to the setting of threshold verifiers when both the verifier and communication are classical.}
\end{theorem}
 
We also generalize Theorem \ref{thm:informal3} to hold for threshold verifiers.
\begin{theorem}[Informal] \label{thm:informal4}
    For any $3$-message protocol with soundness error $s$, the $k$-fold parallel repetition with threshold $0 \leq t \leq k$ has soundness error $\leq f(s)^{O(k)}+\negl(\secp)$, where $f(s) \approx 2^{-\left(t/k-o_k(1)-s\right)^2}$.
\end{theorem}

\begin{remark}
    Our proof is \emph{mildly} non-black-box since we assume knowledge of the cheating prover's size. Our proof is also mildly non-constructive since we use multiple copies of the prover's original state. This latter property is known to be inherent for any reduction that needs to succeed with higher probability than the underlying prover (see for example \cite{BQSY24}). As in \cite{BQSY24}, given a state which is already in a ``good subspace'', we only need one such copy for the reduction to run.
\end{remark}

\paragraph{Open problems.}
    We leave open the question of whether a fully black-box proof can be used to establish parallel repetition in the public-coin setting.\footnote{The proof in \cite{BQSY24} for three-message protocols is black-box in this sense, but it is unclear how to extend it to the multi-round, post-quantum setting.} We also leave open the question of removing the round dependence from our soundness error in Theorems~\ref{thm:informal1} and \ref{thm:informal2}. This was done in the classical setting in \cite{CP15} by analyzing relative entropy; we note that the direct quantum analogue of the classical chain rule does not hold for quantum relative entropy.\footnote{In fact, there are explicit counterexamples to the classical chain rule when directly extended to the setting of quantum channels, and only a modified chain rule is known \cite{FFRS20}.}
    
    Finally, and most interestingly, we leave open the question of obtaining a post-quantum parallel repetition theorem for private-coin protocols. We know that in order to obtain such a result, one would need to modify the underlying protocol, for example by using the random-termination transformation \cite{Hai09}. Our techniques do not seem to extend to the multi-round private-coin setting (see Section \ref{sec:overview} for a discussion).


\section{Technical Overview}\label{sec:overview}
As mentioned in the introduction, the main challenge in the post-quantum (and quantum) setting lies in the fact that most classical reductions for proving parallel repetition inherently assume the ability to perfectly rewind a prover to its previous state. In the setting where the prover's state is quantum, this is impossible to implement efficiently due to the potentially destructive nature of measurement and the no-cloning principle.

Our first approach to bypass this problem is to use a technique introduced in \cite{CMSZ22}, which allows us to rewind the cheating prover to a state that is still  ``useful'' in the sense that it can be used to cheat in the argument system. However, this state is not identical to the original state and the only guarantee we have is that the state is still a ``good'' state in the sense that it is accepting with sufficiently high probability. Unfortunately, this alone turns out not to be enough.  As we elaborate below, the reason is that this ``rewound'' state may record information about past queries used by the reduction, and may deliberately fail on these queries. To make the analysis work, we use a technique called \emph{flooding} \cite{BBK22} which makes non-black box use of the cheating prover. 

Throughout this overview, we focus on the problem of parallel repetition without a threshold (i.e., where the threshold is $t=k$) since this is where our conceptual ideas lie.  We start by explaining our techniques for public-coin protocols.

\subsection{Public-Coin Protocols}\label{sec:overview:public}
We show that for public-coin protocols, parallel repetition reduces the soundness at a (weakly) exponential rate even in the post-quantum setting; this is done by means of a reduction.

We assume that there is a $\QPT$ adversary $\calB$ that succeeds in the $k$-fold repetition setting with probability $\zeta > 0$, and we construct a $\QPT$ adversary $\calA$ which makes black-box calls to $\calB$ and succeeds in a single protocol with probability roughly $1-O\left(m\sqrt{\frac{-\log \zeta}{k}}\right)$, where $m$ is the number of messages sent in the protocol (see Theorem \ref{public_coin_reduction}). This implies that if the original protocol has constant post-quantum soundness then its $k$-fold parallel repetition has post-quantum soundness $\zeta = \exp(-O(k/m^2))$.

\paragraph{The classical setting.} Let us start by recalling the analysis in the classical setting (specifically from the works of \cite{PV07, HPPW10}), and see where it breaks down in the post-quantum setting. In the classical analysis, $\calA$ first picks a random execution $i \in [k]$. The basic idea now is for $\calA$ to forward the verifier's queries to the $k$-fold adversary $\calB$, by embedding in the $i$th execution the query it received from the external verifier, and simulating the queries for all other executions on its own. Namely, in round $j$, upon receiving a query $q_j$ from the external verifier, $\calA$ chooses a random vector $\bar{q}_j$ such that $\bar{q}^i_j = q_j$.\footnote{From now on, we will use superscripts to refer to executions and subscripts to refer to rounds.} Then, when $\calA$ receives answers $\bar{z}_j$ from $\calB$, she checks whether a random continuation of the transcript is accepting. If this continuation is accepting, then she forwards the answer $\bar{z}^i_j$ to the verifier; otherwise she rewinds the prover $\calB$ and tries again with fresh $\bar{q}_j$ such that $\bar{q}_j^i = q_j$. This method of picking the first successful random continuation can be thought of as a noisy measurement of the prover's residual success probability; namely, if the random continuation is successful, then the prover has likely found a ``good'' response.

The classical analysis then compares this reduction (which we label as \textsf{Real}) with an experiment (labeled \textsf{Ideal}) which simulates all verifiers on its own, conditioned on a random continuation being accepting. Raz's lemma (Lemma \ref{raz_lemma}, \cite{Raz95}) is used to argue that these two experiments are close: in particular, one can show that the distribution of $\bar{q}$ ``hides'' the location of the embedded execution. 
Specifically, Raz's lemma asserts that
    \[ \bbE_{i \gets [k], q_j}[\TD((i, (\bar{q}_j \mid W)),(i, (\bar{q}_j \mid W, \bar{q}^i_j = q_j)))] \leq \sqrt{\frac{-\log(\Pr[W])}{k}}, \] 
where $W$ is the event that the random continuation is accepting.

Thus, after $m$ rounds, the expected statistical distance between the two experiments will grow to roughly $m\sqrt{\frac{-\log \zeta}{k}}$, which is why our adversary $\calA$ for the single execution ultimately succeeds with probability $1 - O\left( m\sqrt{\frac{-\log \zeta}{k}} \right)$ (assuming the \textsf{Ideal} experiment succeeds with high probability). We note that the use of Raz's lemma is precisely why our soundness amplification is lossy.\footnote{Indeed, in the classical setting, Chung and Pass \cite{CP15} provide an alternative analysis that avoids this loss by using KL-divergence instead of Raz's lemma. We leave it as an open problem to extend the analysis of \cite{CP15} to the post-quantum setting.}

\paragraph{First challenge in the quantum setting.} The first issue that we are faced with in the post-quantum setting is that perfect rewinding is impossible due to the potentially destructive nature of measurement and the no-cloning principle. We overcome this barrier by utilizing a technique from \cite{CMSZ22} that shows that if we start with a state that has acceptance probability $\zeta$, and we measure only a few bits, say whether the prover's response $\bar{z}_j$ for a query $\bar{q}_j$ is ``good'', then while we cannot rewind back to the state we started with, we can go back to a different state that still has acceptance probability at least $\zeta-\eps$ for any small $\eps$.  The runtime of this ``repairing procedure'' grows polynomially with $1/\eps$. 

\begin{remark}\label{remark:CMSZ}
    The technique of \cite{CMSZ22} was originally designed for the single-round public-coin setting, but can be easily generalized to the multi-round setting. Moreover, it can further be generalized to the private-coin setting, as long as the protocol is computationally simulatable and the verifier's verdict can be efficiently estimated. This latter extension will be useful to obtain parallel repetition for 3-round private-coin protocols.  See Section \ref{sec:overview:private} for details.
\end{remark}
At this point, it is tempting to think that our ability to ``rewind'' the prover's state implies that we can use an analysis similar to that in the classical setting. However, there is still a major obstacle that differentiates the classical setting from the quantum/post-quantum settings. 

\paragraph{Second challenge in the quantum setting.} While in the classical setting, one can rewind the prover back to exactly the same state as before, this is not true in the quantum setting: although one can rewind to a state that has acceptance probability close to $\zeta$, this state may be very different from the original state of the prover. In fact, it may (in theory) crucially remember information about $\bar{q}_j$ and $i$. If it does, then the prover may choose to perform poorly on any input $\bar{q}'_j$ such that $(\bar{q}'_j)^{i} = \bar{q}_j^i = q_j$ (where $q_j$ is the query sent by the external verifier), while at the same time having an acceptance probability close to $\zeta$ on uniformly random queries.

\paragraph{Making the rewound state forgetful.}  

Our strategy will be to force the rewound state to ``forget'' information about $\bar{q}_j$ (and in particular about the external verifier's query) by using a technique due to \cite{BBK22} called \emph{flooding}. The basic idea is to flood the cheating prover $\calB$ with many dummy queries and then argue that $\calB$ cannot remember the embedded query from the external verifier since it does not have enough memory to store all of the queries. The number of dummy queries we need to flood the state with depends on the size of the memory (i.e., number of qubits) that the prover has: if it has $\ell$ qubits, then the number of dummy queries will scale with $\ell$. In particular, this technique is (mildly) non-black-box, and is the only non-black-box part of our reduction. 

We formalize this by presenting a new ``memoryless'' version of the \cite{CMSZ22} lemma, which provides a repairing technique that, while still not rewinding the state to the exact state it started with (which is inherent since performing a measurement can irreversibly destroy information), can rewind the state to one that is almost ``independent'' of the projection applied to the state prior to the rewinding.
In other words, we ensure that the state ``forgets'' the projection. We believe that this lemma (formalized in Lemma~\ref{lem:new-csmz}) is of independent interest and will prove useful for other applications as well. We refer the reader to Section~\ref{sec:new-CMSZ} for details.

\paragraph{The final reduction, outlined.}

Roughly speaking, our final reduction proceeds as follows:
\begin{enumerate}
    \item Estimate the success probability of the $k$-fold cheating prover $\calB$ given an initial state $\ket{\psi}$. With sufficiently (polynomially) many copies of $\ket{\psi}$, one can assume that at least one copy will have a sufficiently good outcome, giving us a residual state in the desired ``good'' subspace. 
    \item Choose a random execution $i \gets [k]$ to embed the external verifier's queries in.
    \item For each round, upon receiving a query from the external verifier, we will try the following sufficiently (polynomially) many times:
    \begin{enumerate}
        \item Generate a $k$-fold query $\bar{q}$ where we embed the external verifier's query in the $i$th coordinate and simulate the rest of the verifiers internally.
        \item Test to see if $\bar{q}$ is a ``good'' query by computing $\calB$'s response to $\bar{q}$ coherently and then running the projection which estimates whether or not the residual success probability of the prover is still high.
        \item
            \begin{enumerate}
                \item If it is, then measure the response, send the answer located in the $i$th coordinate to the external verifier, and proceed to the next round.
                \item Otherwise, we use our memoryless repair procedure to repair the state back to one which is still good over random queries and has almost no memory of the previous query $\bar{q}$.  
            \end{enumerate}
    \end{enumerate}
    \item Conclude when a complete transcript has been formed.
\end{enumerate}
As mentioned above, we instantiate these estimation and memoryless repair procedures using the algorithms given by \cite{CMSZ22}, and tweaking them using ideas from \cite{BBK22} to create ``memoryless'' versions of these procedures. This allows us to test each embedded query many times, since to the prover each test looks nearly indistinguishable from a random query.

\subsection{Private-Coin Protocols}\label{sec:overview:private}

Next, we move to consider the private-coin setting. The first problem that arises is that we can no longer estimate the success probability of a cheating prover since computing the verifier's verdict may require knowing the verifier's private coins. This problem also appears in the classical setting; indeed, it was shown \cite{BIN97} that for private-coin protocols, parallel repetition does not always reduce the soundness error! The issue in the private-coin setting is that the prover may be able to use the verifier's messages from different executions to cheat, since these may actually contain useful information, as opposed to simply consisting of random coins.

\paragraph{Three-message protocols.} Luckily, it turns out that such a counterexample does not exist for the special case of three-message protocols. Indeed, the works of \cite{BIN97}, \cite{CHS05}, and \cite{HPPW10} demonstrated that in the classical setting, all three-message (including private-coin) protocols have exponentially decreasing soundness under parallel repetition. Recently, the work of \cite{BQSY24} extended this result to the quantum (and post-quantum) settings.

In this work, we reprove this result in the post-quantum setting, but through a different framework that allows us to directly generalize to threshold-type verifiers. We take a similar approach to \cite{BIN97} and \cite{HPPW10}, which implement probability estimation via the \emph{soft-decision} approach, which we elaborate on below. We note that \cite{BQSY24} more closely mirrors the \emph{correlation reduction} approach taken by \cite{CHS05}, offering a tight (but more complicated) analysis. Our reduction in the three-message setting follows the same framework as the public-coin reduction with a simple substitution of the probability estimation procedure, again using flooding to make the quantum state ``forgetful'', allowing us to extend our results to the case of threshold-type verifiers.

\begin{remark}
In the three-message setting, one can achieve a straightforward black-box proof, avoiding the need to flood, but at the expense of the reduction being significantly more non-constructive. This is the case both in the post-quantum and quantum settings. Intuitively, one can take a prover that has access to many copies of the original prover's residual state \emph{after a good first response has already been measured}. The downside of this approach is that even if the original prover's starting state was efficiently preparable, its post-measurement residual state is unlikely to be efficiently preparable if the response space is super-polynomially large.
\end{remark}

\paragraph{The soft-decision approach.} Let us recall the soft-decision approach of \cite{BIN97} and \cite{HPPW10}. In the private-coin setting, a major issue is the problem of estimating the success probability of the prover without the external verifier's randomness. A naive first attempt to determining whether a prover's response is good (in the sense that conditioned on this response, the prover's residual success probability is relatively high) might be to assume the external coordinate is good if all other coordinates are good as well. However, this intuitive approach does not quite work: consider the following example first given by Chung \cite{Chu11} of a malicious prover $\calB$ for the $k$-parallel-repeated protocol that will foil such a reduction. Suppose $\calB$ interacts with the parallel verifier in such a way that it is accepted by all executions with probability $\delta^k$, and for each $j \in [k]$, is accepted by all executions except the $j$th execution with probability $(1-\delta^k)/k$. Intuitively, conditioned on all but the $i$th coordinate accepting, $\calB$ may actually only succeed on the $i$th coordinate with probability $\frac{\delta^k}{\delta^k+(1-\delta)^k/k} \ll \delta$. Such a prover can be considered to be one that has ``bad correlations''.

In order to deal with provers with this behavior, classical analyses have generally taken two approaches, the correlation reduction approach of \cite{CHS05} (which is roughly the approach taken by \cite{BQSY24}) and the soft-decision approach of \cite{BIN97} and \cite{HPPW10} (which is the approach we take in this work). Instead of taking only responses for which \emph{all} executions, except the external execution, accept, we weigh our decision to take a response with probability exponentially proportional to the number of rejecting executions outside the external execution. That is, if $z$ executions excluding the embedded one are rejecting, then we will accept the response with probability roughly $2^{-z}$. Looking at the prover in our example, it is easy to see that this soft-decision approach will choose answers that are rejected on coordinate $i$ only with probability $O(1/k)$. In general, this intuition explains how the soft-decision approach works to ``smoothen'' the distribution of errors made by the reduction. This approach also almost immediately generalizes to the setting of threshold verifiers, where we can set our acceptance probability to be roughly $\min\{1, 2^{k-t-z}\}$, where $t$ is the threshold of the parallel verifier.\footnote{We note that in the classical setting, the soft-decision approach is known to be inherently lossy compared to the correlation reduction approach.}

\paragraph{Challenges in the quantum setting.}

The first challenge, which we encountered in the public-coin setting as well, is that perfect rewinding cannot be done in the quantum setting. As in the public-coin setting, we use \cite{CMSZ22} to repair. As mentioned above, we cannot rewind to the exact same state, and we can only guarantee that we repair to a state that is still successful, i.e., succeeds on a random continuation with sufficiently high probability. As mentioned in Remark \ref{remark:CMSZ}, the technique of \cite{CMSZ22}, which  was originally designed for the single-round public-coin setting, can be easily extended to multiple rounds, as long as the protocol is computationally simulatable and the verifier's verdict can be efficiently estimated. Here, we run into an issue which is common to the classical setting as well -- given the prefix of a transcript, how can we estimate if a continuation is accepting without access to the external verifier's randomness?

In the three-message setting, this is not a problem, since if the prefix consists of only the first prover message, then one can perfectly simulate a full transcript and compute the verifier's verdict, and if the prefix consists of a full transcript then one can implement the soft-decision approach to estimate the prover's success probability (see Remark \ref{remark:CMSZ_queries} for more details).

The fact that this rewinding is not perfect also brings with it challenges which we first encountered in the public-coin setting. Specifically, the state can depend on its projection history, and in particular can remember the query from the external verifier, and fail on that specific query while succeeding on random queries. In the public-coin setting we solved this issue by flooding. In fact, one can observe that flooding can always be applied to the first two messages of the protocol, which suffices for the three-message setting. The reason is that whenever we flood, we need to be able to sample the external verifier's query (and its private randomness) from the conditional distribution given the transcript prefix, but for the first two messages this prefix is empty, so in this case we can efficiently sample from the randomness distribution.

\paragraph{Multi-round protocols.} Unfortunately, in the multi-round, private-coin setting (where the number of messages sent is more than three) we are unable to prove a parallel repetition theorem in the post-quantum setting. The first obstacle that arises is that after the first verifier's message is sent, it may be hard to simulate the verifier's subsequent messages without access to the private randomness of the verifier. This is a challenge that occurred even in the classical setting (where random termination enabled the overcoming of this problem). 

In the post-quantum setting, even if the protocol is computationally simulatable, we still do not know how to prove a parallel repetition theorem. The problem is that every time we estimate the success probability of the prover we must do so without knowing the embedded execution's final verdict, and in the post-quantum setting, the state may remember that the embedded execution, denoted by $i$, is not being correctly estimated, and once it remembers $i$, it can behave in a way so that only the $i$'th execution (which has a fixed query given by the external verifier) will be rejected. 

With both public-coin and three-message protocols, the issue of the state remembering $i$ arose, but we were able to make the state ``forgetful'' by flooding. In the public-coin case, simulation of messages and verdict is trivial, and in the three-message case, simulation of the messages and verdict is possible since there is only one verifier message. However, in the multi-round setting, it is apparent that we cannot flood, since we cannot simulate the messages of the verifier and its verdict, after the first message has been fixed.

Indeed, our proofs in the public-coin and three-message settings rely substantially on the ability to flood, and this inability to flood in the general private-coin setting is one major obstacle to proving a parallel repetition theorem with the ``estimate and repair'' approach. 

\paragraph{A note on flooding.} We note that the key property required to apply the flooding technique is the ability to simulate the messages and the verdict of the external verifier given a prefix of the transcript. This allows us to show similar parallel repetition results for protocols which are public-coin except for (possibly) the \emph{last} round, which includes both the public-coin and three-message settings. As many protocols for classical verification of quantum computation are public-coin except for the \emph{first} round, determining if such protocols have exponentially decreasing soundness under parallel repetition seems to be one potential extension of interest.\footnote{We note that even a restriction to such protocols with only four messages would already have useful implications.}

\section{Preliminaries}\label{sec:preliminaries}
We use $\PPT$ as a shorthand for probabilistic polynomial time and $\QPT$  as a shorthand for quantum polynomial time.

\begin{notation}
    For any two quantum states $\brho_1$ and $\brho_2$, we use the notations
    \[
    \brho_1\stackrel{\eps}\equiv\brho_2 ~~\mbox{ and }~~\TD(\brho_1,\brho_2)\leq\eps
    \]
    to denote that the trace distance between $\brho_1$ and $\brho_2$ is at most $\eps$. In particular, this means that for any quantum adversary $\calA$,
    \[
    |\Pr[\calA(\brho_1)=1]-\Pr[\calA(\brho_2)=1]|\leq \eps.
    \]
\end{notation}

\begin{definition}[Measurement]
    A \emph{real-valued measurement} $\textsf{M}$ is a quantum circuit that takes as input a quantum state $\brho$ and outputs a quantum state $\brho^*$ and a real number $p^*$.
\end{definition}

\begin{definition}[$(\eps, \delta)$-Almost Projective Measurement, \cite{CMSZ22}]
    A real-valued measurement $\textsf{M}$ on a Hilbert space $\mathcal{H}$ is said to be \emph{$(\eps, \delta)$-almost projective} if on input a quantum state $\brho\in S(\mathcal{H})$, $\sfM$ satisfies 
    \begin{align*}
            \Pr\left[|p^{*}-p^{**}| \geq \eps \hspace{3pt} \middle| \hspace{5pt}
            \begin{aligned}
                &(\boldsymbol{\rho}^{*}, p^{*}) \leftarrow \sfM(\boldsymbol{\rho}) \\
                &(\boldsymbol{\rho^{**}}, p^{**}) \leftarrow \sfM(\boldsymbol{\rho^{*}})
            \end{aligned}\right] \leq \delta.
        \end{align*}
\end{definition}

We next define a family of measurements that are $(\eps, \delta)$-almost projective for arbitrary $\eps,\delta>0$.
\begin{definition}[Almost Projective Measurement] \label{def:almost-projection}
    A real-valued measurement family $\textsf{M}=\{\textsf{M}_{\eps,\delta}\}_{\eps,\delta>0}$ on $\mathcal{H}$ is said to be \emph{almost projective} if for all quantum states $\brho\in S(\mathcal{H})$ and all parameters $\eps, \eps', \delta, \delta' \in (0, 1]$,
    \begin{align*}
        \Pr\left[|p^{*}-p^{**}| \geq \max\{\eps, \eps'\} \hspace{3pt} \middle| \hspace{5pt}
        \begin{aligned}
            &(\boldsymbol{\rho}^{*}, p^{*}) \leftarrow \sfM_ {\eps, \delta}(\boldsymbol{\rho}) \\
            &(\boldsymbol{\rho^{**}}, p^{**}) \leftarrow \sfM_{\eps', \delta'}(\boldsymbol{\rho^{*}})
        \end{aligned}\right] \leq \max\{\delta, \delta'\}.
    \end{align*}
    In particular, for all $\eps, \delta > 0$, $\sfM_{\eps, \delta}$ is an $(\eps, \delta)$-almost projective measurement.
\end{definition}

We will use the well-known fact that any quantum measurement performed by a quantum circuit can be implemented projectively by deferring measurement. That is, any quantum circuit $C$ can be equivalently implemented by applying a unitary dilation $U$ (corresponding to replacing intermediate measurements with CNOT gates) followed by a measurement of the output register and an application of $U^{\dagger}$.

\subsection{Interactive Protocols with Computational Soundness}\label{prelim:IA}

In what follows, we focus on interactive protocols between two parties, where one party, denoted by $V$ (for verifier), outputs a single bit, indicating acceptance or rejection. The other party,  denoted by $P$ (for prover), tries to convince the verifier to accept. We define the notion of computational soundness for such protocols. For the sake of simplicity, we focus on the case where the parties receive as input the security parameter $1^\secp$, and omit all additional inputs. We denote by
\[ \langle{(P,V)(1^\secp)\rangle} \]
the output bit of $V$ after interacting with $P$.

\begin{definition}
An interactive protocol between parties $P$ and $V$, where $V$ outputs a single bit is said to have (computational) soundness $\eps$ for any polynomial-size cheating prover ${\cal A}$ there exists a negligible function $\mu$ such that for every $\secp\in\mathbb{N}$
    \[
    \Pr[\langle{(\calA,V)(1^\secp)\rangle}=1]\leq \eps+\mu(\secp).
    \]
\end{definition}

Interactive protocols with computational soundness are also referred to as interactive arguments, and we will use these terms interchangeably. A public-coin interactive argument is a protocol where the verifier's messages to the prover consist entirely of random coins, and the verifier's verdict is deterministic and publicly computable given a complete transcript. We consider only protocols where the communication between the prover and verifier is classical and the verifier is a classical $\PPT$ algorithm.

\begin{remark}
    We assume without loss of generality that the last message of any protocol is sent from the prover to the verifier. We call an interactive protocol an $m$-round protocol if it consists of $m$ rounds of back and forth messages between the prover and verifier (as opposed to the alternative $2m$-round definition). For the sake of simplicity and without loss of generality, we also assume that the first message is sent by the verifier (unless we are in the three-message setting), as otherwise the verifier can always begin by sending coins that are ignored in the rest of the protocol.
\end{remark}

\begin{notation}
    The transcript of an $m$-round interactive protocol $(P, V)$ consists of all messages sent between the prover and verifier, denoted by $\tau = (q_1, z_1, \dots, q_m, z_m)$, where $q_j$ is the query sent by~$V$ in the $j$th round, and $z_j$ is the response sent by~$P$ in the $j$th round. In the case of three-message protocols, we instead use the notation $\tau = (z_1, q, z_2)$ to emphasize that only one query is sent by the verifier and that the first message is sent by the prover.
\end{notation}  


\subsection{Value Estimation}

\noindent We borrow the following lemma from \cite{CMSZ22} (using the formalism from \cite{LMS22}).
\begin{lemma}[\textbf{Value Estimation and State Repair}, \cite{CMSZ22,LMS22}] \label{CMSZ}
    Let $\mathcal{H}$ be a Hilbert space. There exist quantum algorithms:
    \begin{enumerate}
        \item $(\boldsymbol{\rho^{*}}, p^{*}) \leftarrow \ValEst_{V,A}(\boldsymbol{\rho}, \eps, \delta)$ is given black-box access to a verifier circuit $V: \{0, 1\}^d \times \{0, 1\}^n \to \{0, 1\}$ and a quantum circuit $A$, and on input a quantum state $\boldsymbol{\rho} \in S(\mathcal{H})$ and accuracy parameters $\eps, \delta \in (0, 1]$, outputs a quantum state $\boldsymbol{\rho^{*}} \in S(\mathcal{H})$ and value $p^{*} \in X \subseteq [-\frac{1}{2}, \frac{3}{2}]$ for some discrete set $X$ where $N_{\eps, \delta} := |X| = O\left(\frac{1}{\eps^2} \log \frac{1}{\delta}\right)$.\footnote{The reader should think of $p^{*}$ as a noisy estimate of the success probability, whose expectation is in $[0,1]$.}
        \item $\boldsymbol{\sigma^{*}} \leftarrow \Repair_{\sfM, \Pi}(\boldsymbol{\sigma}, y, p, \eta)$ is given black-box access to an $(\eps, \delta)$-almost projective measurement $\sfM$ and a projective measurement $\Pi=(\Pi_y)_{y \in Y}$ on $\mathcal{H}$, and on input a quantum state $\boldsymbol{\sigma} \in S(\mathcal{H})$, an outcome $y \in Y$, a probability $p \in [0, 1]$, and parameter $\eta \in (0, 1]$, outputs a quantum state $\boldsymbol{\sigma^{*}} \in S(\mathcal{H})$.
    \end{enumerate}
    These algorithms satisfy the following guarantees:
    \begin{enumerate}
        \item \textbf{Value Estimation:}
        \begin{align*}
            \mathbb{E}\left[p^{*} \hspace{3pt} \middle| \hspace{5pt} (\boldsymbol{\rho^{*}}, p^{*}) \leftarrow \ValEst_{V,A}(\boldsymbol{\rho}, \eps, \delta)\right] = \emph{Pr}\left[V(r, z) = 1 \hspace{3pt} \middle| \hspace{5pt} \begin{aligned}
                &r \leftarrow \{0, 1\}^d \\
                &z \leftarrow A(\boldsymbol{\rho}, r)
            \end{aligned}\right].
        \end{align*}
        \item \textbf{$\ValEst^{V, A}_{\eps, \delta} := \ValEst_{V, A}(\cdot, \eps, \delta)$ is an Almost Projective Family:}
        \begin{align*}
            \Pr\left[|p^{*}-p^{**}| \geq \max\{\eps, \eps'\} \hspace{3pt} \middle| \hspace{5pt}
            \begin{aligned}
                &(\boldsymbol{\rho}^{*}, p^{*}) \leftarrow \ValEst_{V, A}(\boldsymbol{\rho}, \eps, \delta) \\
                &(\boldsymbol{\rho^{**}}, p^{**}) \leftarrow \ValEst_{V, A}(\boldsymbol{\rho^{*}}, \eps', \delta')
            \end{aligned}\right] \leq \max\{\delta, \delta'\}.
        \end{align*}
        \item \textbf{Repairing:}
        For any $(\eps, \delta)$-almost projective measurement $\sfM$ on $\calH$,
        \begin{align*}
            \emph{Pr}\left[|p^{*}-p^{**}| \geq 2\eps \hspace{3pt} \middle| \hspace{5pt} \begin{aligned}
                &(\boldsymbol{\rho^{*}}, p^{*}) \leftarrow \sfM(\boldsymbol{\rho}) \\
                &(\boldsymbol{\sigma}, y) \leftarrow \Pi(\boldsymbol{\rho^{*}}) \\
                &\boldsymbol{\sigma^{*}} \leftarrow \Repair_{\sfM, \Pi}(\boldsymbol{\sigma}, y, p^{*}, \eta) \\
                &(\boldsymbol{\rho^{**}}, p^{**}) \leftarrow \sfM(\boldsymbol{\sigma^{*}})
            \end{aligned}\right] \leq N \cdot (\eta+\delta) + 4\sqrt{\delta},
        \end{align*}
        where $N=|Y|$ is the number of outcomes the projective measurement $\Pi$ can obtain.
    \end{enumerate}
    $\ValEst$ and $\Repair$ are oracle circuits with $O(|X|)=O\left(\frac{\log 1/\delta}{\eps^2}\right)$ and $O(1/\eta)$ gates, respectively.
\end{lemma}

\begin{remark}[Computing $\ValEst$ coherently]
We note that by deferring measurements, we can coherently compute $\ValEst$ over some register using ancilla registers.
\end{remark}

\begin{remark}[Randomness in $\ValEst$]
Note that even when the prover strategy has a fixed success probability of precisely 0 or 1, the $\ValEst$ procedure may alter the prover's state. Thus, in the case where the circuits $A$ and $V$ are trivial (i.e. when $d = n = 0$), we define $\ValEst$ to be the \emph{deterministic} algorithm which simply returns the output of $V$ and does not modify the quantum state. This extension/modification of $\ValEst$ clearly still satisfies properties 1 and 2.
\end{remark}

\begin{remark}[Computing queries in $\ValEst$]\label{remark:CMSZ_queries}
While the $\ValEst$ procedure was originally defined in the context of single-round, public-coin protocols, we can easily generalize it to the multi-round setting (as was done in \cite{LMS22}) 
by having the randomness of $V$ specify all of the verifier's randomness in the interactive protocol. With some slight modifications, one can also extend $\ValEst$ to work in the private-coin (three-message) setting by considering a restricted adversary which separately computes the verifier's query before applying a specified unitary corresponding to the prover's strategy (see Remark \ref{remark:computing_private_queries} for more details). This allows us to efficiently simulate random executions and estimate the prover's success probability.
\end{remark}

\subsection{Information Theory} \label{sec:info}

\begin{lemma}[\cite{BBK22}]\label{flooding}
    Let $\Vec{Y} = (Y_1, \ldots, Y_t)$ be a joint distribution over $t$ classical random variables. Let $\Vec{y}$ be distributed according to $\Vec{Y}$. Let $\boldsymbol{s}$ be an $\ell$-qubit random variable that has arbitrary dependence on $\Vec{y}$. We let $\Vec{y}_i$ denote the prefix $\Vec{y}_i = (y_1, \ldots, y_i)$ for $1 \leq i \leq t$, and let $\Vec{y}_0$ be the empty vector (and likewise for $\Vec{Y}$). Let $J$ be the uniform distribution over $[t]$ and let $j \leftarrow J$. Define $y' \leftarrow Y_j | (\Vec{Y}_{j-1} = \Vec{y}_{j-1})$. Then it holds that 
    \[\TD((j, \Vec{y}_{j-1}, y_j, \boldsymbol{s}), (j, \Vec{y}_{j-1}, y', \boldsymbol{s})) \leq \sqrt{\ell/(2t)}.\]
\end{lemma}

\begin{lemma}[Raz's Lemma, \cite{Raz95, Hol07, Hai09}]\label{raz_lemma}
    Let $X_1, \ldots, X_k$ be independent random variables over some probability space, and let $\overbar{X} = (X_1, \ldots, X_k)$. Let $W$ be a non-empty event on the same space. Then
    \[ \bbE_{i\leftarrow [k]}[\TD((i, (\overbar{X} | W)),(i, (\overbar{X} | W, X_i)))] \leq \sqrt{-\log(\Pr[W])/k}. \]
\end{lemma}

\section{A Memoryless CMSZ}\label{sec:new-CMSZ}
The work of \cite{CMSZ22} (Lemma~\ref{CMSZ}) provides a technique for running a quantum adversary $\calA$ interacting with a verifier $V$, measuring its response $z$, and then repairing its state to one that is still accepted by $V$ with roughly the same probability as the original state, where the runtime of the repair procedure grows polynomially with the number of possible prover responses. This is a powerful technique that allows us to rewind in the post-quantum setting. However, this rewinding is different from the rewinding in the classical setting in that the  repaired state may be significantly different from the original state, and in particular may contain information about queries on which the adversary $\calA$ was run. This makes it problematic to apply this technique to lift reductions to the post-quantum setting, and for parallel repetition in particular. 

Specifically, in the analysis of parallel repetition, we reduce a cheating prover $\calB$ for the $k$-fold repetition to a cheating prover $\calA$ for a single execution. In a nutshell, $\calA$ interacts with a verifier, it receives a query $q$, embeds this query in a random execution $i \gets [k]$, and chooses the queries for the rest of the executions on its own. It sends these $k$ queries to $\calB$, computes the response of $\calB$ (coherently) and estimates its success probability. If the success probability is not sufficiently high then it applies the $\Repair$ procedure and does this again with fresh queries in all executions except the embedded one. The issue is that the repaired state may remember information about the $k$ queries on which it was run, and eventually remember the embedded query $q$, and be repaired to a state that generates accepted answers with high probability but always generates a rejected answer on the embedded query $q$. See Section~\ref{sec:overview:public} for a detailed discussion.   

In this section we show how to make this $\Repair$ procedure {\em forgetful}, so that the repaired state does not have information about the embedded query. This requires the forgetful $\Repair$ procedure to have a bound on the size of the quantum state of $\calA$, and thus is (mildly) non-black-box.

\subsection{Distributional Memoryless CMSZ}

The idea for making the $\Repair$ procedure from Lemma~\ref{CMSZ} forgetful is to run the adversary, followed by running the $\Repair$ procedure, sufficiently many times on dummy queries, so that the state will not be able to remember all of these queries due to lack of space. If the distribution of the dummy queries is the same as the distribution of the real query then the state should not be able to distinguish the real query from the dummy ones, and in this way, we control the information that can be remembered about the real query by the number of dummy queries we flood with.

We formalize this by defining two algorithms: $\Prepare$ and $\Repair'$ (to distinguish it from $\Repair$ from Lemma~\ref{CMSZ}). 
The algorithm $\Prepare$ starts the flooding process. It is given black-box access to the family of projective measurements, denoted by $\calP$, as well as access to an almost projective family $\sfM=\{\sfM_{\eps,\delta}\}$. Intuitively, $\Prepare$ repeats the following ``sufficiently many times,'' where the number of times is a random variable that depends on the size of the quantum state (denoted by $\ell$).
\begin{enumerate}
    \item Apply~$\sfM$ to the state to obtain a value $p'$.
    \item Choose at random a projection $\Pi\gets \calP$ and apply $\Pi$ to the state.
    \item Apply $\Repair$ to the state, which repairs the state back to a state for which $\sfM$ would give a value close to $p'$ with high probability.
\end{enumerate} 
The $\Repair'$ algorithm first applies $\Repair$ to the state and then continues the flooding process, exactly as was done by $\Prepare$ in steps (1)-(3) above.  The guarantees are formalized in the following lemma.

\begin{lemma}\label{lem:new-csmz}
    Let $\mathcal{H}$ be a Hilbert space. There exist quantum algorithms:
    \begin{enumerate}
        \item $(\bfrho', p') \gets \Prepare_{\sfM, \calP}(\bfrho, \eps, \delta, \eta, \ell)$ is given black-box access to an almost projective measurement family $\sfM$ and to a family of projective measurements $\calP$ associated with a set of $N$ outcomes in $Y=\{y_1,\ldots,y_N\}$. It takes as input a state $\bfrho \in S(\mathcal{H})$ and parameters $\eps, \delta, \eta \in (0, 1]$ and $\ell \in \bbN$, and outputs a quantum state $\bfrho'$ and a value $p'$.
        \item $\bsigma^{*} \leftarrow \Repair'_{\sfM, \calP, \Pi}(\bsigma, y, p, \eps, \delta, \eta, \ell)$ is given black-box access to $\sfM$ and $\calP$ as above as well as black-box access to a projection $\Pi \in \calP$. It takes as input a quantum state $\bsigma \in S(\mathcal{H})$, an outcome $y \in Y$, a value $p$, and parameters $\eps, \delta, \eta \in (0, 1]$ and $\ell \in \bbN$, and outputs a quantum state $\bsigma^{*} \in S(\mathcal{H})$.
    \end{enumerate}
    These algorithms satisfy the following guarantees:
    \begin{enumerate}
        \item \textbf{Size:}
        $\Prepare$ and $\Repair'$ are oracle circuits of size $\poly(\ell, 1/\eta)$ where all oracle calls to $\sfM$ are with parameters $\eps'$ and $\delta'$ that are polynomially related to $\eps,\delta,\eta,1/\ell$.
        \item \textbf{Preparing Preserves Functionality:}
        For every almost-projective measurement family $\sfM$ and every family of $N$-outcome projections $\calP$,
        \begin{align*}
            \emph{Pr}\left[|p^{*}-p'| \geq 4\eps 
            \hspace{3pt} \middle| \hspace{5pt} \begin{aligned}
                &(\bfrho^{*}, p^{*}) \leftarrow \sfM_{\eps, \delta}(\bfrho) \\
                &(\bfrho', p') \gets \Prepare_{\sfM, \calP}(\bfrho^*, \eps, \delta, \eta, \ell)
            \end{aligned}\right] \leq N \cdot (\eta + 4\delta).
        \end{align*}
        In addition, 
        \begin{align*}
            \emph{Pr}\left[|p'-p^{**}| \geq \eps 
            \hspace{3pt} \middle| \hspace{5pt} \begin{aligned}
                &(\bfrho^{*}, p^{*}) \leftarrow \sfM_{\eps, \delta}(\bfrho) \\
                &(\bfrho', p') \gets \Prepare_{\sfM, \calP}(\bfrho^{*}, \eps, \delta, \eta, \ell) \\
                &(\bfrho^{**}, p^{**}) \leftarrow \sfM_{\eps, \delta}(\bfrho')
            \end{aligned}\right] \leq \delta.
        \end{align*}
        \item \textbf{Repairing is Functional but Forgetful:} 
        For every almost-projective measurement family $\sfM$, every family of $N$-outcome projections $\calP$, and every $\Pi\in\calP$,
        \begin{align*}
            \emph{Pr}\left[|p^{*}-p^{**}| \geq 4\eps \hspace{3pt} \middle| \hspace{5pt} \begin{aligned}
                &(\bfrho^{*}, p^{*}) \leftarrow \sfM_{\eps, \delta}(\bfrho) \\
                &(\bfrho', p')\gets \Prepare_{\sfM, \calP}(\bfrho^*, \eps, \delta, \eta, \ell)\\
                &(\bsigma, y) \leftarrow \Pi(\bfrho') \\
                &\bsigma^{*} \leftarrow \Repair'_{\sfM, \calP, \Pi}(\bsigma, y, p', \eps, \delta, \eta, \ell) \\
                &(\bfrho^{**}, p^{**}) \leftarrow \sfM_{\eps, \delta}(\bsigma^{*})
            \end{aligned}\right] \leq N \cdot (\eta + 4\delta).
        \end{align*}
        Moreover, for every $\sfM, \calP$ as above, and for every $\ell$-qubit state $\bfrho$ it holds that 
        \[ \TD((\Pi, \bsigma^{*}_{\Pi}), (\Pi', \bsigma^{*}_{\Pi})) \leq N \cdot \eta \]
        where $\Pi,\Pi'\leftarrow\calP$ and where the state $\bsigma^{*}_{\Pi}$ is generated as follows:  
        \begin{enumerate}
            \item Compute $(\bfrho^{*}, p^{*}) \leftarrow \sfM_{\eps, \delta}(\bfrho)$.
            \item Compute $(\bfrho', p')\gets \Prepare_{\sfM,\cal P}(\bfrho^*, \eps, \delta, \eta, \ell)$.
            \item Compute $(\bsigma, y) \leftarrow\Pi(\bfrho')$.
            \item Compute $\bsigma^{*}_{\Pi} \leftarrow \Repair'_{\sfM, \calP, \Pi}(\bsigma, y, p', \eps, \delta, \eta, \ell)$.
            \item Output $\bsigma^{*}_{\Pi}$.
        \end{enumerate} 
    \end{enumerate}
\end{lemma}

\begin{proof}
    Let us define $\Prepare$ and $\Repair'$, where $\Repair$ is the procedure defined in Lemma \ref{CMSZ}:
    \begin{definition}[$\Prepare_{\sfM, \calP}$]
    On input $(\bfrho^{*}, \eps, \delta, \eta, \ell)$:
        \begin{enumerate}
            \item Set $\bsigma'' = \bfrho^{*}$ and $T = 4\ell/\eta^3$. Sample $t \leftarrow [T]$ uniformly. Define $\sfM' := \sfM_{\eps/2T, \delta^2/64T^2}$.
            \item For $i = 1$ to $t-1$: 
            \begin{enumerate}
                \item Compute $(\bfrho', p') \leftarrow \sfM'(\bsigma'')$.
                \item Sample $\Pi_i \leftarrow \calP$ uniformly and compute $(\bsigma', y) \leftarrow \Pi_i(\bfrho')$.
                \item Compute $\bsigma'' \leftarrow \Repair_{\sfM', \Pi_i}(\bsigma', y, p', \eta/2T)$.
            \end{enumerate}
            \item Compute $(\bfrho', p') \leftarrow \sfM'(\bsigma'')$.
            \item Return $(\bfrho', p')$.
        \end{enumerate}
    \end{definition}
    \begin{definition}[$\Repair'_{\sfM, \calP, \Pi}$]
    On input $(\bsigma, y, p', \eps, \delta, \eta, \ell)$:
        \begin{enumerate}
            \item Set $T = 4\ell/\eta^3$. Define $\sfM' := \sfM_{\eps/2T, \delta^2/64T^2}$.
            \item Compute $\bsigma^{*} \leftarrow \Repair_{\sfM', \Pi}(\bsigma, y, p', \eta/2T)$.
            \item For $i = 1$ to $T$: 
            \begin{enumerate}
                \item Compute $(\bfrho, p^{**}) \leftarrow \sfM'(\bsigma^{*})$.
                \item Sample $\Pi_i \leftarrow \calP$ uniformly and compute $(\bsigma, y) \leftarrow \Pi_i(\bfrho)$.
                \item Compute $\bsigma^{*} \leftarrow \Repair_{\sfM', \Pi_i}(\bsigma, y, p^{**}, \eta/2T)$.
            \end{enumerate}
            \item Compute $(\bsigma^{*}, p^{**}) \leftarrow \sfM'(\bsigma^{*})$.
            \item Return $\bsigma^{*}$.
        \end{enumerate}
    \end{definition}
    We begin by analyzing the size of $\Prepare$ and $\Repair'$. First, note that both procedures have oracle access to $\sfM$ and $\calP$, and $\Repair'$ has oracle access to $\Pi$, and all oracle calls to $\sfM$ are with parameters $\eps/2T = \Theta(\eps \eta^3/\ell)$ and $\delta^2/64T^2 = \Theta(\delta \eta^6/\ell^2)$, which are polynomially related to $\eps, \delta, \eta, 1/\ell$, as desired.
    
    In $\Prepare$, at most $T$ calls are made to $\sfM'$ and to random $\Pi_i\gets\calP$, outside of $\Repair$. Additionally, $\Repair$ makes $O(T/\eta)$ calls to $\sfM'$ and $\Pi_i$ (for each $i \leq t-1$) for a total of $O(T) + O(T^2/\eta)$ calls to measurements in $\sfM$ and $\calP$. Thus, $\Prepare$ is an oracle circuit of size $O(T) + O(T^2/\eta) = O(\ell/\eta^3)+O(\ell^2/\eta^7) = \poly(\ell, 1/\eta)$.

    Similarly, $\Repair'$  makes $T$ calls to $\sfM'$ and random $\Pi_i\gets\calP$, outside of $\Repair$, and its $O(T)$ calls to $\Repair$ themselves make $O(T/\eta)$ calls to $\sfM'$ and $\Pi_i$ (for each $i \leq t-1$). Thus, $\Repair'$ is also an oracle circuit of size $O(T) + O(T^2/\eta) = O(\ell/\eta^3)+O(\ell^2/\eta^7) = \poly(\ell, 1/\eta)$.
    
    Next, we prove the first part of property 3. Let us fix any sampling of $t \in [T]$ in the first step of $\Prepare$ and prove our claim for this fixed $t$. Let us denote by $p_0$ the output of the first $\sfM_{\eps, \delta}$, denote by $p_i$ the output of the $\sfM' := \sfM_{\eps/2T, \delta^2/64T^2}$ measurement in the $i$th loop of $\Prepare$, and denote by $p_t$ the output of the final $\sfM'$ call in $\Prepare$. For each $j \in [T]$ let $p_{t+j}$ denote the output of the $j$th call to $\sfM'$ in $\Repair'$, and $p_{t+T+1}$ denote the output of the final $\sfM'$ call in $\Repair'$. Let $p_{t+T+2}$ denote the output of the second $\sfM_{\eps, \delta}$ call outside of $\Prepare$ and $\Repair'$.

    Showing the first half of property 3 is equivalent to proving that 
    \[\Pr[|p_0 - p_{t+T+2}| \geq 4\eps] \leq N \cdot (\eta+4\delta).\]
    Observing that $\delta \leq 1$ and $T = 4 \ell / \eta^3 \geq 1$, we know that $\eps/T \leq \eps$ and $\delta^2/64T^2 \leq \delta$. Thus, it follows that
        \[ \Pr[|p_0 - p_1| \geq \eps], \Pr[|p_{t+T+1} - p_{t+T+2}| \geq \eps] \leq \delta \]
    by definition, and for all $1 \leq i \leq t+T$,
        \[ \Pr\left[|p_{i} - p_{i+1}| \geq \frac{\eps}{T}\right] \leq N \cdot \left(\frac{\eta}{2T} + \frac{\delta^2}{64T^2}\right)+4\sqrt{\frac{\delta^2}{64T^2}} \]
    by Property 3 of Lemma \ref{CMSZ}.
    
    Next, let 
    \[\mathsf{Good} := (|p_0-p_1| \leq \eps) \land (|p_{t+T+1}-p_{t+T+2}| \leq \eps) \land \left(|p_i-p_{i+1}| \leq \frac{\eps}{T}\right)_{i=1}^{t+T}\] 
    denote the event that all probabilities are close. Since $t + T \leq 2T$, it follows by a union bound that
    \begin{align*}
        \Pr[\mathsf{Good}] &\geq 1-\left[\delta+\delta+N \cdot \left(\frac{\eta}{2T} + \frac{\delta^2}{64T^2}\right) \cdot (t+T) + 4\sqrt{\frac{\delta^2}{64T^2}} \cdot (t+T)\right] \\
        &\geq 1-\left[\delta+\delta+N \cdot \left(\frac{\eta}{2T} + \frac{\delta^2}{64T^2}\right) \cdot 2T + 4\sqrt{\frac{\delta^2}{64T^2}} \cdot 2T\right] \\
        &= 1-[2\delta + N \cdot \eta + N \cdot \delta^2/32T + \delta] \\
        &\geq 1 - [3\delta + N\cdot \eta + N \cdot \delta] \\
        &\geq 1-N\cdot(\eta+4\delta).
    \end{align*}
    Now note that if $\mathsf{Good}$ occurs, then 
    \begin{align*}
        |p_0 - p_{t+T+2}| \leq \sum_{i=0}^{t+T+1}|p_i - p_{i+1}| \leq \eps+\eps+\frac{\eps}{T}\cdot (t+T) \leq 4\eps.
    \end{align*}
    Thus, 
    \[\Pr[|p^{*}-p^{**}| \geq 4\eps] \leq \Pr[\overbar{\mathsf{Good}}] \leq N \cdot (\eta+4\delta),\]
    as desired.

    We note that the first half of property 2 is equivalent to stating that
        \[ \Pr[|p_0-p_t| \geq 4\eps] \leq N \cdot (\eta+4\delta), \]
    which follows from the same argument as above. The second half of property 2 follows directly by observing that the last step of $\Prepare$ (which determines $p'$) is a measurement $\sfM'$ and applying property 2 of Lemma \ref{CMSZ}.

    Finally, to prove the second half of property 3, we first fix the state $\bfrho$ and consider the current implementation:
    \begin{mdframed}
        \textbf{Hybrid $\mathcal{H}_1$:}
        \begin{enumerate}
            \item Sample $\Pi \leftarrow \calP$.
            \item Compute $(\bfrho^{*}, p^{*}) \leftarrow \sfM_{\eps, \delta}(\bfrho)$.
            {\color{red} \item Compute $(\bfrho', p')\gets \Prepare_{\sfM, \cal P}(\bfrho^*, \eps, \delta, \eta, \ell)$.
            \item Compute $(\bsigma, y) \leftarrow\Pi(\bfrho')$.
            \item Compute $\bsigma^{*}_{\Pi} \leftarrow \Repair'_{\sfM, \calP, \Pi}(\bsigma, y, p', \eps, \delta, \eta, \ell)$.}
            \item Output $(\Pi, \bsigma^{*}_{\Pi})$.
        \end{enumerate}
    \end{mdframed}
    In what follows, we open the definition of $\Prepare$ and $\Repair'$, and replace steps 3-5 to produce the following identical distribution:
    \begin{mdframed}
        \textbf{Hybrid $\mathcal{H}_2$:}
        \begin{enumerate}
            \item Sample $\Pi \leftarrow \calP$.
            \item Compute $(\bfrho^{*}, p^{*}) \leftarrow \sfM_{\eps, \delta}(\bfrho)$.
            {\color{red} \item Set $\bsigma'' = \bfrho^{*}$  and $T = 4\ell/\eta^3$. Sample $\Pi_1, \ldots, \Pi_T \leftarrow \calP$ uniformly. Sample $t \leftarrow [T]$ uniformly and \emph{replace} $\Pi_{t}$ with $\Pi$. Define $\sfM' := \sfM_{\eps/2T, \delta^2/64T^2}$.
            \item For $i = 1$ to $T$: 
                \begin{enumerate}
                    \item Compute $(\bfrho', p') \leftarrow \sfM'(\bsigma'')$.
                    \item Compute $(\bsigma', y_{\Pi_i}) \leftarrow \Pi_i(\bfrho')$.
                    \item Compute $\bsigma'' \leftarrow \Repair_{\sfM', \Pi_i}(\bsigma', y_{\Pi_i}, p', \eta/2T)$.
                \end{enumerate}
            \item For $i = T+1$ to $T+t$: 
            \begin{enumerate}
                \item Compute $(\bfrho', p') \leftarrow \sfM'(\bsigma'')$.
                \item Sample $\Pi_i \leftarrow \calP$ uniformly and compute $(\bsigma', y_{\Pi_i}) \leftarrow \Pi_i(\bfrho')$.
                \item Compute $\bsigma'' \leftarrow \Repair_{\sfM', \Pi_i}(\bsigma', y_{\Pi_i}, p', \eta/2T)$.
            \end{enumerate}
            \item Compute $(\bsigma^{*}, p^{**}) \leftarrow \sfM'(\bsigma'')$.}
            \item Output $(\Pi, \bsigma^{*})$.
        \end{enumerate}
    \end{mdframed}
    Next, we note that since $t$, $\Pi$, and $\Pi_{t}$ are sampled uniformly randomly, we can move the sampling of $t$ to after step 4 and drop the planting of $\Pi$ while maintaining an identical output distribution:
    \begin{mdframed}
        \textbf{Hybrid $\mathcal{H}_3$:}
        \begin{enumerate}
            \item Compute $(\bfrho^{*}, p^{*}) \leftarrow \sfM_{\eps, \delta}(\bfrho)$.
            \item Set $\bsigma'' = \bfrho^{*}$ and $T = 4\ell/\eta^3$. Sample $\Pi_1, \ldots, \Pi_T \leftarrow \calP$ uniformly. Define $\sfM' := \sfM_{\eps/2T, \delta^2/64T^2}$.
            \item For $i = 1$ to $T$: 
                \begin{enumerate}
                    \item Compute $(\bfrho', p') \leftarrow \sfM'(\bsigma'')$.
                    \item Compute $(\bsigma', y_{\Pi_i}) \leftarrow \Pi_i(\bfrho')$.
                    \item Compute $\bsigma'' \leftarrow \Repair_{\sfM', \Pi_i}(\bsigma', y_{\Pi_i}, p', \eta/2T)$.
                \end{enumerate}
            \item Sample $t \leftarrow [T]$ uniformly and set $\Pi = \Pi_{t}$.
            \item For $i = T+1$ to $T+t$: 
            \begin{enumerate}
                \item Compute $(\bfrho', p') \leftarrow \sfM'(\bsigma'')$.
                \item Sample $\Pi_i \leftarrow \calP$ uniformly and compute $(\bsigma', y_{\Pi_i}) \leftarrow \Pi_i(\bfrho')$.
                \item Compute $\bsigma'' \leftarrow \Repair_{\sfM', \Pi_i}(\bsigma', y_{\Pi_i}, p', \eta/2T)$.
            \end{enumerate}
            \item Compute $(\bsigma^{*}, p^{**}) \leftarrow \sfM'(\bsigma'')$.
            \item Output $(\Pi_t, \bsigma^{*})$.
        \end{enumerate} 
    \end{mdframed}
    Note that steps 5 and 6 of $\calH_3$ have \emph{no dependence} on $\Pi = \Pi_{t}$, so it suffices to show that the state $\bsigma$ at the end of step 4 satisfies 
    \[ \TD((\Pi_t, \bsigma), (\Pi', \bsigma)) \leq N \cdot \eta \]
    for uniformly sampled $\Pi' \leftarrow \calP$.
    
    To this end, for every $j\in[N]$, define the random variable $C_j := |\{i\in[T]: y_{\Pi_i} = y_j\}|$ denoting the number of projections with outcome equal to $y_j$ in steps 1-3, where $Y = \{y_1, \ldots, y_N\}$ is the set of $N$ possible outcomes for all projections in $\calP$. Note that for every $j \in [N]$,
    \begin{align*}
        \Pr[(y_{\Pi_{t}} = y_j) \land (C_j \leq \eta T/2)] &=\Pr[y_{\Pi_{t}} = y_j \mid C_j \leq \eta T/2] \cdot \Pr[C_j \leq \eta T/2] \\
        &\leq \Pr[y_{\Pi_{t}} = y_j \mid C_j \leq \eta T/2] \leq \eta/2,
    \end{align*}
    where the probability is over the randomness in $\calH_3$.
    Thus, by a union bound we have that
    \[\Pr[\exists j \in [N]: (y_{\Pi_{t}} = y_j) \land (C_j \leq \eta T/2)] \leq N \cdot \eta/2. \]
    Let $B$ be the event that there exists $j \in [N]$ such that $y_{\Pi_{t}} = y_j$ and $C_j \leq \eta T/2$. By a simple coupling argument, we have that
    \begin{align*}
    \TD((\Pi_{t}, \bsigma), (\Pi', \bsigma)) &= \Pr[B] \cdot \TD((\Pi_{t}, \bsigma \mid B), (\Pi', \bsigma \mid B)) + \Pr[\overbar{B}] \cdot \TD\left(\left(\Pi_{t}, \bsigma \mid \overbar{B}\right), \left(\Pi', \bsigma \mid \overbar{B}\right)\right) \\
    &\leq N \cdot \eta/2 + \TD\left(\left(\Pi_{t}, \bsigma \mid \overbar{B}\right), \left(\Pi', \bsigma \mid \overbar{B}\right)\right).
    \end{align*}
    But now note that the event $\overbar{B}$ implies that for \emph{every} possible outcome of $y_{\Pi_{t}}$, at least $\frac{\eta}{2}$ fraction of the $T$ projections result in the same outcome as $y_{\Pi_{t}}$. Recall that all $\Pi \leftarrow \calP$ are drawn from an identical distribution, so all projections which observed $y_{\Pi_{t}}$ are drawn from the \emph{same} distribution as $\Pi_{t}$. Thus,  by Lemma \ref{flooding},
    \begin{align*}
        \TD\left(\left(\Pi_{t}, \bsigma \mid \overbar{B}\right), \left(\Pi', \bsigma \mid \overbar{B}\right)\right) \leq \sqrt{\frac{\ell}{2(\frac{\eta}{2} \cdot T)}} = \eta/2,
    \end{align*}
    and so we conclude that
    \[ \TD((\Pi, \bsigma), (\Pi', \bsigma)) := \underset{t \leftarrow [T]}{\TD}((\Pi_{t},\bsigma), (\Pi', \bsigma)) \leq N \cdot \eta/2 + \eta/2 \leq N \cdot \eta, \]
    as desired.
\end{proof}

\begin{remark}[Efficiency]
    We observe that when $V$ and $A$ are efficiently computable circuits (in some security parameter, say $\lambda$), then $\ValEst^{V, A}_{\cdot, \cdot}$ is an almost projective family such that for any $\eps, \delta$, $\ValEst^{V, A}_{\eps, \delta}$ runs in time $\poly(\lambda, 1/\eps, \log 1/\delta)$. If we then set $\sfM = \ValEst^{V, A}_{\cdot, \cdot}$ and assume $\calP$ is an efficiently implementable projection family (i.e. it runs in $\poly(\lambda, 1/\eps, \log 1/\delta)$ time) we see that both $\Prepare$ and $\Repair'$ can run in time $\poly(\lambda, 1/\eps, \log 1/\delta, 1/\eta, \ell)$ with \emph{no} oracle calls.
\end{remark}

\section{Parallel Repetition for Public-Coin Protocols}\label{sec:public-coin}
In this section, we analyze the soundness of $k$-fold parallel repetition of public coin protocols, and prove that the soundness goes down exponentially in $k$. 
For any interactive protocol $(P,V)$ we denote by $(P^{(k)},V^{(t, k)})$ its $t$-threshold, $k$-fold parallel repetition.  Namely, $V^{(t, k)}$ accepts at the end of the protocol if and only if at least $t$ out of its $k$ constituent verifiers accept.
\subsection{Our Results}
We prove the following parallel repetition theorem for public-coin protocols:
\begin{theorem}\label{public_coin_reduction} 
    Let $(P, V)$ be a $m$-round (where $m = m(\secp)$ is polynomial) public-coin interactive argument. Suppose that there exists a quantum adversary $\calB$, $\xi=\xi(\secp) > 0$, and polynomials $1 \leq t = t(\secp) \leq k=k(\secp)$ such that for all $\secp\in \mathbb{N}$,
    \begin{equation}\label{public_coin_parallel_cheater}
        \Pr[\langle (\calB, V^{(t, k)})(1^\secp) \rangle = 1] \geq \xi.
    \end{equation}
    Then there exists a quantum adversary $\calA$ running in time $\poly(|\calB|, \secp, 1/\xi)$ and a negligible function $\negl$ such that for every $\secp\in \mathbb{N}$,
    \[
        \Pr[\langle (\calA, V)(1^\secp) \rangle = 1] \geq \frac{t}{k}-2m\sqrt{\frac{-\log(\xi/3m^2)}{k}}-\negl(\secp).
    \]
\end{theorem}

Note that Theorem \ref{public_coin_reduction} implies the following corollary:
\begin{corollary}\label{public_coin_parallel_repetition} 
    Let $(P, V)$ be a $m$-round (where $m = m(\secp)$ is polynomial) public-coin interactive argument and let $\eps = \eps(\secp)$ be any parameter such that for every $\QPT$ adversary $\calA$ there exists a negligible function $\negl$ such that for every $\secp \in \mathbb{N}$,
    \[
        \Pr[\langle (\calA, V)(1^\secp) \rangle = 1] \leq \eps(\secp)+\negl(\secp).
    \]
    Fix any polynomials $1 \leq t = t(\secp) \leq k = k(\secp)$ such that $\eps < t/k$.\footnote{The condition that $\eps < t/k$ is necessary as otherwise any adversary $\calA$ which succeeds with probability at least $\eps$ can be easily converted (by a direct product) into a $k$-fold adversary which already convinces $\geq \eps k \geq t$ verifiers on expectation (and hence succeeds with probability at least 1/2).} Then for every $\QPT$ adversary $\calB$ there exists a negligible function $\negl$ such that for every $\secp\in\mathbb{N}$,
    \[
        \Pr[\langle (\calB, V^{(t, k)})(1^\secp) \rangle = 1] \leq \max\{6m^2 \cdot \exp\left(\frac{-k}{4m^2} \cdot \left(\frac{t}{k}-\eps\right)^2\right), \negl(\secp)\}.
    \]
\end{corollary}
\begin{proof}[Proof of Corollary \ref{public_coin_parallel_repetition}]
   Suppose for the sake of contradiction that there exists a $\QPT$ adversary $\calB$ that succeeds against $V^{(t, k)}$ with non-negligible probability $\xi > 6m^2 \cdot \exp\left(\frac{-k}{4m^2} \cdot \left(\frac{t}{k}-\eps\right)^2\right) > 0$. Applying Theorem \ref{public_coin_reduction} gives a quantum adversary $\calA$ running in time $\poly(\secp, 1/\xi)$ such that
    \begin{align*}
        \Pr[\langle (\calA, V)(1^\secp) \rangle = 1] &\geq \frac{t}{k}-2m\sqrt{\frac{-\log(\left[6m^2 \cdot \exp\left(\frac{-k}{4m^2} \cdot \left(\frac{t}{k}-\eps\right)^2\right)\right]/3m^2)}{k}}-\negl(\secp) \\
        &= \frac{t}{k}-2m\sqrt{\frac{-\log(2\exp\left(\frac{-k}{4m^2} \cdot \left(\frac{t}{k}-\eps\right)^2\right))}{k}}-\negl(\secp) \\
        &= \frac{t}{k}-2m\sqrt{\frac{-\log(\exp\left(\frac{-k}{4m^2} \cdot \left(\frac{t}{k}-\eps\right)^2\right))-1}{k}}-\negl(\secp) \\
        &= \frac{t}{k}-2m\sqrt{\frac{\frac{k}{4m^2} \cdot \left(\frac{t}{k}-\eps\right)^2-1}{k}}-\negl(\secp) \\
        &= \frac{t}{k}-\sqrt{\left(\frac{t}{k}-\eps\right)^2-\frac{4m^2}{k}}-\negl(\secp) \\
        &\geq \frac{t}{k}-\left[\left(\frac{t}{k}-\eps\right)-\frac{\frac{4m^2}{k}}{2\left(\frac{t}{k}-\eps\right)}\right]-\negl(\secp) \labeleq\label{inequality}\\
        &\geq \frac{t}{k}-\left(\frac{t}{k}-\eps\right)+\frac{\frac{4m^2}{k}}{2\left(\frac{t}{k}-\eps\right)}-\negl(\secp) \\
        &= \eps + \frac{\frac{4m^2}{k}}{2\left(\frac{t}{k}-\eps\right)}-\negl(\secp) \\
        &\geq \eps + \frac{\frac{4m^2}{k}}{\frac{2t}{k}}-\negl(\secp) \\
        &= \eps + \frac{2m^2}{t}-\negl(\secp),
    \end{align*}
    where \eqref{inequality} follows from the fact that $\sqrt{x^2-y} \leq x-\frac{y}{2x}$ for $x, y > 0$ such that $x^2 - y > 0$\footnote{The condition holds here since $c := 6m^2 \cdot \exp\left(\frac{-k}{4m^2} \cdot \left(\frac{t}{k}-\eps\right)^2\right) < \xi \leq 1$, and so $2m\sqrt{\frac{-\log(c/3m^2)}{k}} = \sqrt{(\frac{t}{k}-\eps)^2-\frac{4m^2}{k}} > 0$.}. Since $m$ and $t$ are polynomial, $\eps+\frac{2m^2}{t}-\negl(\secp) > \eps+\negl(\secp)$.
    
    But since $\calB$ has non-negligible advantage, then $1/\xi$ is polynomial in $\secp$ (infinitely often) and so $\calA$ is $\QPT$ (infinitely often), in contradiction to the original assumption that no such $\QPT$ adversary exists. Thus, no such $\QPT$ adversary $\calB$ exists either.
\end{proof}

\begin{proof}[Proof of Theorem \ref{public_coin_reduction}]

We assume that $m \geq 2$, since parallel repetition in the case of $m = 1$ can be easily shown using a test-and-check reduction (one can repeatedly test queries on fresh states until the first acceptance).

Denote by $\ket{\psi}$ the initial state of $\calB$, and let $\Lambda = \poly(\secp)$ be the number of qubits in $\ket{\psi}$. For the sake of simplicity, we assume without loss of generality that $\ket{\psi}$ contains designated registers $\calZ_1, \calZ_2, \dots, \calZ_m$, where $\calZ_j$ is the register on which $\calB$ writes its response to the $j$'th round queries. We denote all the other registers in $\ket{\psi}$ by $\calI$. We write $Q_j^{(k)}$ to refer to the set of possible queries sent in round $j$ in the $k$-fold repeated protocol $(P^{(k)},V^{(t, k)})$ and $\bar{q}^i_j$ to refer to the $i$th coordinate of a query vector $\bar{q}_j \in Q_j^{(k)}$.

For every $j \in [m]$ we denote by $U_j$ the collection of unitaries indexed by $j$th round queries $\bar{q}_j$ such that $U_j(\bar{q}_j)$ is the unitary that $\calB$ applies to its quantum state upon receiving query $\bar{q}_j$ (which operates on $\calI$ and $\calZ_j$).  We denote by 
\begin{equation}\label{eqn:Uj} 
    U^{(j)}(\bar{q}_{j}, \bar{q}_{j+1}, \dots, \bar{q}_m) = U_m(\bar{q}_m) U_{m-1}(\bar{q}_{m-1}) \dots U_{j}(\bar{q}_{j}). 
\end{equation}
The assumption that each $U_j(\bar{q}_{j})$ is a unitary is without loss of generality, as any adversary not of this form can be ``purified'' into an adversary with unitary strategies, identical observable behavior, and constant factor blowup in size. 

For a complete transcript $\tau_m$ and index $i \in [k]$, define $\Accept_i(\tau_m) = 1$ iff the output of the $i$th verifier at the end of a protocol with transcript $\tau_m$ is `1', and $\Accept(\tau_m) = 1$ iff $\sum_i \Accept_i(\tau_m) \geq t$. For any partial transcript $\tau = (\bar{q}_1, \bar{z}_1, \dots,\bar{q}_{j-1}, \bar{z}_{j-1})$, we define a classical function $V_\tau$ that takes as input a continuation of the transcript, namely $(\bar{q}_j,\bar{z}_j, \bar{q}_{j+1}, \bar{z}_{j+1}, \dots, \bar{q}_m, \bar{z}_m)$, and outputs the (threshold) parallel verifier's decision predicate 
    \[ V^{(t, k)}(\bar{q}_1, \bar{z}_1, \dots, \bar{q}_m, \bar{z}_m) := \Accept(\bar{q}_1 || \bar{z}_1 || \dots || \bar{q}_m || \bar{z}_m). \]
Additionally, for a partial transcript $\tau = (\bar{q}_1, \bar{z}_1, \dots,\bar{q}_{\ell-1}, \bar{z}_{\ell-1})$, query $\bar{q}_\ell$, and parameters $\ell, \eps, \delta$, denote by $U_{\ValEst}(\tau, \bar{q}_\ell, \ell, \eps, \delta)$ the unitary that first coherently computes $\bar{z}_\ell = U_\ell(\bar{q}_\ell)$ and then coherently computes $\ValEst_{V_{\tau || \bar{q}_\ell || \bar{z}_\ell}, U^{(\ell+1)}}(\cdot, \eps, \delta)$ (writing the outcome of the $\ValEst$ procedure in the ancilla register $\calE$ using the ancilla register $\calF$). We think of $U^{(\ell+1)}$ as a quantum circuit that takes as input $(\bar{q}_{\ell+1},\ldots,\bar{q}_m)$ and outputs  corresponding answers $(\bar{z}_{\ell+1},\ldots,\bar{z}_m)$ (according to $U^{(\ell+1)}$), and we think of $V_{\tau || \bar{q}_\ell || \bar{z}_\ell}$ as the circuit that generates random queries $(\bar{q}_{\ell+1},\ldots,\bar{q}_m)$\footnote{Since the protocol is public-coin, $\bar{q}_{\ell+1}, \ldots, \bar{q}_m$ are independent/non-adaptively chosen random coins and can thus be sampled concurrently. Therefore, given access to $U^{(\ell+1)}$, a random continuation of the protocol can be thought of as a single-round quantum game. Hence, $\ValEst$ can be implemented essentially without modification (as in \cite{LMS22}).} and upon receiving answers $(\bar{z}_{\ell+1},\ldots,\bar{z}_m)$ outputs $\Accept(\tau_m)$ for $\tau_m \triangleq (\tau || \bar{q}_\ell || \bar{z}_\ell || \bar{q}_{\ell+1} || \bar{z}_{\ell+1}||\ldots|| \bar{q}_m || \bar{z}_m)$. 
    
\paragraph{Informal description of the reduction $\calA$.} Before we give the full reduction and analysis, we first informally describe our proof strategy. The adversary $\calA$ uses as subroutines the algorithms $\ValEst$, $\Prepare$, and $\Repair'$ as defined in Lemmas \ref{CMSZ} and \ref{lem:new-csmz}, and does the following: 
\begin{enumerate}
    \item Apply $\ValEst$ to the state $\ket{\psi}$ (which is the initial state of $\calB$), to obtain a probability $p_0$. Given sufficiently many copies of the state $\ket{\psi}$, with high probability, for at least one of these states $p_0$ will be close to $\xi$, which is the success probability of $\calB$. Specifically, we can assume that $p_0 \geq \xi-\eps_0$ for some small $\eps_0$ (which we set to be $\eps_0:=\frac{\xi}{m^2})$.
    \item Choose a random execution $i \gets [k]$ to embed the queries in.
    \item For each round $\ell \in [m]$, upon receiving a query $q_\ell$ from the external verifier $V$, do the following $\iter = \poly(\secp, m, 1/\xi)$ times:
    \begin{enumerate}
        \item Generate a $k$-fold query $\bar{q}_\ell$ such that its $i$'th coordinate is equal to $q_\ell$ (i.e., embed the external verifier's query in the $i$th execution and simulate the remaining $k-1$ verifiers internally). Namely, generate  $\bar{q}_\ell\gets Q_{\ell}^{(k)}$ such that $\bar{q}_\ell^i=q_\ell$.
        \item Let $\Pi=\CheckCoins_{\bar{q}_\ell}$ be the projection that estimates the acceptance probability of the answers given by $\calB$ on queries $\bar{q}_\ell$. We would like to apply $\Pi$ to the state so that if the resulting probability is not sufficiently high (close to $\xi$), then we can repair it in a \emph{forgetful} way so that $\bar{q}_\ell$ (and in particular $\bar{q}^i_\ell$) is forgotten. This is implemented as instructed by Lemma~\ref{lem:new-csmz}, as follows:
     
        Let $V_{\tau_{\ell-1}}$ be the verifier circuit which samples queries $(\bar{q}_\ell,\ldots,\bar{q}_m)$, and upon receiving responses $(\bar{z}_\ell,\ldots,\bar{z}_m)$, outputs its verdict $\Accept(\tau_{\ell-1},\bar{q}_\ell,\bar{z}_\ell,\ldots,\bar{q}_m,\bar{z}_m)$.  In what follows, for ease of presentation, we slightly abuse notation and also denote by $U^{(\ell)}$ the adversary circuit which upon receiving queries $(\bar{q}_\ell,\ldots,\bar{q}_m)$ applies the unitary $U^{(\ell)}(\bar{q}_\ell,\ldots,\bar{q}_m)$ (which describes the answers of $\calB$ from round $\ell$ onward, conditioned on the transcript in the first $\ell-1$ rounds being $\tau_{\ell-1}$; see Equation~\eqref{eqn:Uj}) to its state, thereby producing responses $\bar{z}_\ell, \ldots, \bar{z}_m$ in registers $\calZ_{\ell}, \ldots, \calZ_m$.\footnote{Note that we also use $U^{(\ell)}$ to refer to the same unitary which is implemented by this circuit.} We can now define $M^{\ell}=\ValEst_{V_{\tau_{\ell-1}},U^{(\ell)}}$, which is an almost-projective measurement family.
        
        \begin{enumerate}
            \item \label{item:first} Apply $M^\ell$ to the state to obtain a probability estimate $p_{\ell-1}$.
            \item Run $\Prepare$ with respect to the family of projections $\calP=\{\CheckCoins_{\bar{q}_\ell}\}_{{\bar{q}_\ell}\in Q_\ell^{(k)}}$ and the family of almost projections~$M^\ell$, on input the relevant parameters, including $p_{\ell-1}$.
            \item Apply $\Pi$ to the state to obtain a probability estimate~$p$.
            \item If $p$ is not sufficiently high (i.e. $p < \xi-\ell\epsilon_0$) then repair the state by applying the (forgetful) $\Repair'$ procedure to it, and try again (go back to Item~\ref{item:first}).
            \item If $p$ is sufficiently high (i.e. $p\geq \xi-\ell\epsilon_0$) then measure the answer register $\calZ_\ell$ to obtain answers  $\bar{z}_\ell$ corresponding to $\bar{q}_\ell$, send $\bar{z}^i_\ell$ to the verifier, and proceed to round $\ell+1$. 
        \end{enumerate}
    \end{enumerate}
\end{enumerate}

We now formally describe algorithm $\calA$ in the Figure \ref{alg:A-public-desc} below.
\defbox{Projection}{$\CheckCoins_{\ell, \tau_{\ell-1}, \bar{q}}[\bfrho]$}{CheckCoins}{
    Let $L$ be the number of ancilla qubits required to compute $U_{\ValEst}(\tau_{\ell-1}, \bar{q}, \ell, \eps, \delta)$.
    \begin{enumerate}
        \item
            Apply $U_{\ValEst}(\tau_{\ell-1}, \bar{q}, \ell, \eps, \delta)$ to $\bfrho \otimes \ket{0^L}\bra{0^L}_{\calE, \calF}$, resulting in state $\sum_{z_\ell, p_{z_{\ell}}} (\bfrho'_{z_\ell})_{\calI, \calZ_{-\ell}} \otimes \ket{z_\ell}\bra{z_\ell}_{\calZ_\ell} \otimes \ket{p_{z_\ell}}\bra{p_{z_\ell}}_{\calE} \otimes \ket{\mathsf{junk}_{z_\ell, p_{z_{\ell}}}}\bra{\mathsf{junk}_{z_\ell, p_{z_{\ell}}}}_{\calF}$.
        \item
            Measure the bits in register $\calE$, obtaining $p$ and remaining state $\bfrho''$. 
        \item
            Apply $U_{\ValEst}(\tau_{\ell-1}, \bar{q}, \ell, \eps, \delta)^{\dagger}$ to get state $\bsigma = (\bfrho''')_{\calI, \calZ} \otimes \ket{0^L}\bra{0^L}_{\calE, \calF}$. Discard registers $\calE$ and $\calF$ before returning $(p, \bfrho''')$.
    \end{enumerate}
}

\algo{The Algorithm $\calA$}{A-public-desc}{    
    Let $\iter := \secp m^2/\xi$ and $\Lambda := \mathsf{size}(\ket{\psi})$ be the number of qubits of $\ket{\psi}$. $\calA$ begins with state $\ket{\psi}^{\otimes \iter}$, and runs as follows:
    
    \begin{enumerate}
    \item
        Set $\eps_0 := \frac{\xi}{m^2}, \eps := \frac{\eps_0}{16 \cdot \iter}, \delta := \min\{2^{-\secp}, 2^{-k}\}, \eta := \frac{1}{2km \cdot \iter}/N_{\eps, \delta}$, where $N_{\eps, \delta}$ is the parameter from Lemma \ref{CMSZ}. Sample $i \gets [k]$. Define $\tau_0 := \bot$. 
    \item
        For $s = 1$ to $\iter$:
        \begin{enumerate}
            \item[(a)]
                Set $\bfrho = \ket{\psi}$ to be $\calA$'s $s$th copy of $\ket{\psi}$. Run $(p_0, \bfrho_0) \gets \ValEst_{V_{\tau_0}, U^{(1)}}(\bfrho, \eps, \delta)$.
                \begin{enumerate}
                    \item
                        If $p_0 \geq \xi - \eps_0$, continue to step 3.
                    \item
                        Else, if $p_0 < \xi - \eps_0$ and $s = \iter$, return $(i, \bot)$.
                \end{enumerate}
        \end{enumerate}
    \item
        For $\ell = 1$ to $m$: 
        \begin{enumerate}
            \item[(a)]
                Let $\calP := \{\CheckCoins_{\ell, \tau_{\ell-1}, \bar{q}}\}_{\bar{q} \in Q_{\ell}^{(k)}}$ be the projection family defined above. Define the almost projective measurement family $\sfM^{\ell} := \{\ValEst_{V_{\tau_{\ell-1}}, U^{(\ell)}}(\cdot, \eps, \delta)\}_{\eps, \delta > 0}$.
            \item[(b)]
                Receive the $\ell$th-round message $q_{\ell}$ from $V$. 
            \item[(c)] For $s = 1$ to $\iter$:
            \begin{enumerate}
                \item Generate a query $\bar{q}_{\ell} \gets Q_{\ell}^{(k)}$ uniformly conditioned on $\bar{q}^i_{\ell} = q_\ell$. Let $\Pi := \CheckCoins_{\ell,\tau_{\ell-1}, \bar{q}_{\ell}}$. 
                \item Compute $(\bfrho_{\ell-1}', p_{\ell-1}) \gets \sfM^{\ell}_{\eps, \delta}(\bfrho_{\ell-1})$.
                \item Compute $(\bfrho_{\ell-1}'', p'_{\ell-1}) \gets \Prepare_{\sfM^{\ell}, \calP}(\bfrho_{\ell-1}', p_{\ell-1}, \eps, \delta, \eta, \Lambda)$. 
                
                If $p'_{\ell-1} < \xi - \ell \eps_0 - (4s+1) \eps$, return $(i, \bot)$.
                \item Compute $(\bsigma, p) \gets \Pi[\bfrho_{\ell-1}'']$.
                
                If $p \geq \xi-(\ell+1)\eps_0$, skip to step 3(d). 
                
                Else, if $p < \xi-(\ell+1)\eps_0$ and $s = \iter$, return $(i, \bot)$.
                \item Compute $\bfrho_{\ell-1} = \Repair'_{\sfM^{\ell}, \calP, \Pi}(\bsigma, p, p_{\ell-1}', \eps, \delta, \eta, \Lambda)$.
            \end{enumerate}
            \item[(d)]
                Apply $U_{\ValEst}(\tau_{\ell-1}, \bar{q}_{\ell}, \ell, \eps, \delta)$ to $\bsigma \otimes \ket{0^L}\bra{0^L}_{\calE, \calF}$ before measuring the $\calZ_{\ell}$ register to get response $\bar{z}_\ell$. Discard/measure the ancilla registers $\calE$ and $\calF$, resulting in collapsed state $\bsigma^{*}$. Set $\bfrho_{\ell} = \bsigma^{*}$.
            \item[(e)]
               Set $\tau_{\ell} = \tau_{\ell-1} || \bar{q}_{\ell} || \bar{z}_{\ell}$ and send $\bar{z}^i_{\ell}$ back to $V$. 
        \end{enumerate}
    \item
        Return $(i, \tau_m)$.
    \end{enumerate}
}
\noindent
\textbf{Runtime.}
The fact that $\calA$ runs in time $\poly(|\calB|, \secp, 1/\xi)$ follows from the efficiency of $\ValEst'$, $\Prepare$, and $\Repair'$, and the fact that all parameters are polynomial in $\secp$ and $1/\xi$.
$ $\newline
\textbf{Soundness.}
Before we analyze the soundness of our reduction, we note that we can assume without loss of generality that $\xi \gg \delta$, as otherwise our guarantee is essentially trivially true.

\noindent We begin by considering the following set of hybrids $\calH_j$ defined for each $j \in \{0, \ldots, m\}$:
\begin{itemize}
    \item Hybrid $\calH_j$: For rounds $1 \leq \ell \leq j$, replace the sampling of $\bar{q}_{\ell}$ in step 3(c)(i) with a completely uniform sample of $\bar{q}_{\ell} \gets Q^{(k)}_{\ell}$ and derive partial transcript $\tau_j = (\bar{q}_1, \bar{z}_1, \ldots, \bar{q}_j, \bar{z}_j)$. Now, set $i \gets [k]$ (instead of at step 1) and emulate a random execution of $(\calA, V)$ conditioned on $(i, \tau_j)$ being the transcript of the protocol after the first $j$ rounds to derive a full transcript $\tau_m$ and state $\bfrho_m$. Return $(i, \tau_m)$.
\end{itemize}
Note that by definition, $\calH_0$ produces the distribution of outputs over a random execution of $(\calA, V)$.

Before we continue, we note a simple fact about delaying measurements:
\begin{fact}\label{fact:deferring_measurement}
    Fix any state $\brho$, partial transcript $\tau=(\bar{q}_1,z_1,\ldots,\bar{q}_{\ell-1},z_{\ell-1})$, and $\ell$'th round query $\bar{q}_\ell$. Then the following two procedures return identical output distributions:
    \begin{itemize}
        \item Apply $U_{\ValEst}(\tau, \bar{q}_\ell, \ell, \eps, \delta)$ to the state $\brho\otimes\ket{0^L}\bra{0^L}_{\calE, \calF}$, where $\calE, \calF$ are the ancilla registers that are initialized to $0^L$ (where $L$ denotes the number of ancilla qubits used by $U_{\ValEst}$). Then measure registers $\calE$, followed by measuring registers $\calZ_{\ell}$ and $\calF$, to get $p, z_{\ell}, \mathsf{junk}$, and a collapsed state $\bsigma$. Return $(z_{\ell}, p, \bsigma)$, where $\bsigma$ excludes the $\calE$ and $\calF$ registers that were measured and discarded.
        \item Apply $U_{\ell}(\bar{q}_\ell)$ to $\brho$ and measure register $\calZ_{\ell}$ to get $z_{\ell}$ and state $\brho^{*}$. Apply 
            \[ (p, \bsigma) \gets \ValEst_{V_{\tau || \bar{q}_\ell || z_\ell}, U^{(\ell+1)}}(\bfrho'_{z_{\ell}}, \eps, \delta)(\brho^{*}) \]
       and return $(z_{\ell}, p, \bsigma)$.
    \end{itemize}
\end{fact}
The above fact follows from the principle of deferred measurement, which asserts that we can always delay the measurement of both $p$ and $z_{\ell}$ by introducing ancilla registers $\calF$ which are initialized to $0^L$. Once all measurements are deferred, the order of measurement clearly does not matter.

In what follows we first prove that $\calH_m$ returns an accepting transcript with high probability, and then use a hybrid argument to argue that this implies that $\calH_0$ returns an accepting transcript with high probability.
Proving that $\calH_m$ returns an accepting transcript with high probability is done in two steps. First, we prove that the probability that $\calH_m$ returns an accepting transcript is precisely the probability that it does not abort, and then we prove that the probability that it aborts is small.
\begin{lemma}\label{public_abort}
    $\underset{(i, \tau_m) \gets \calH_m}{\Pr} [\Accept(\tau_m) = 1] = 1-\underset{(i, \tau_m) \gets \calH_m}{\Pr}[\tau_m = \bot].$
\end{lemma}
\begin{proof}[Proof of Lemma \ref{public_abort}]
It suffices to prove that if $\calH_m$ does \emph{not} abort and returns a transcript, then this transcript must be accepting. If $\calH_m$ does not abort, then this means that in round $\ell = m$, there is some final $\bar{q}^{*}$ such that $\CheckCoins_{m, \tau_{m-1}, \bar{q}^{*}}$ on some state $\brho_{m-1}$ returned $p \geq \xi - (m+1)\eps_0$. Note that step 3 of $\CheckCoins$ is undone by step 3(d) of $\calH_m$. This means that we applied $U_{\ValEst}$, then measured some $p^{*} \geq \xi - (m+1)\eps_0$, followed by measuring the response $\bar{z}_m$, and $\mathsf{junk}_{z_\ell}$. By Fact \ref{fact:deferring_measurement}, this is equivalent to having measured $\bar{z}_m$ and computing $\ValEst$ on the resulting \emph{full} transcript and receiving some probability $p^{*} \geq \xi - (m+1) \eps_0 = \xi - \frac{m+1}{m^2} \xi > 0$. But when $\ell = m$ this $\ValEst$ is the \emph{deterministic} function which computes precisely whether the full transcript is accepting and must either have value 0 or 1, and so $p^{*} = 1$. Thus, by our setting of parameters, the full transcript is accepting as long as $\calH_m$ does not abort.
\end{proof}

We next argue that $\calH_m$ aborts only with small probability.
\begin{lemma}\label{final_public_success}
    $\underset{(i, \tau_m) \gets \calH_m}{\Pr} [\tau_m = \bot] \leq 1/k+\negl(\secp).$
\end{lemma}
\begin{proof}[Proof of Lemma \ref{final_public_success}]
We first bound the probability that $\calH_m$ aborts in step 2:
\begin{claim}\label{public_step_two}
    $\calH_m$ aborts in step 2 with probability at most $\negl(\secp)$.
\end{claim}
\begin{proof}[Proof of Claim \ref{public_step_two}]
    Since the starting state $\ket{\psi}$ has success probability which is at least $\xi$, by property 1 of Lemma \ref{CMSZ}, we have that 
    \[ \bbE[p_0 | (p_0, \bfrho_0) \gets \ValEst_{V_{\tau_0}, U^{(1)}}(\ket{\psi}, \eps, \delta)] \geq \xi. \]
    Recall that since $\ValEst$ takes on values in $[-\frac{1}{2}, \frac{3}{2}]$ (see Lemma \ref{CMSZ}), by a simple Markov argument, the probability that a given copy of $\ket{\psi}$ will succeed in step 2(a) is
    \begin{align*}
        \Pr\left[p_0 \geq \xi - \eps_0 \hspace{3pt} \middle| \hspace{5pt} (p_0, \bfrho_0) \gets \ValEst_{V_{\tau_0}, U^{(1)}}(\ket{\psi}, \eps, \delta)\right] \geq 2\eps_0/3.
    \end{align*}
    Thus, by using a Chernoff bound we conclude that the probability of failing to succeed with any of the $\iter = \secp/\eps_0$ copies (and thus aborting in step 2) is at most $\negl(\secp)$.
\end{proof}
To bound the probability of aborting in step 3, we will rely on the following basic fact from probability theory, that for every set of events $\{E_i\}_{i=1}^n$ it holds that 
\[
    \Pr[\vee_i E_i] = \sum_i \Pr[E_i \mid (\wedge_{j < i} \overbar{E_j})].
\]
Thus, it suffices to bound the conditional probability of each ``bad'' event occurring assuming all previous ``bad'' events have not occurred.

\begin{claim}\label{public_step_three}
    Conditioned on not aborting in step 2, $\calH_m$ aborts in step 3 with probability at most $\frac{1}{k}+\negl(\secp)$.
\end{claim}
\begin{proof}[Proof of Claim \ref{public_step_three}]
    We begin by analyzing step 3(c)(ii) when $s = 1$ and distinguish between the case where $\ell = 1$ and $\ell > 1$:
    \begin{itemize}
        \item $\ell = 1$: the $\ValEst$ measurement falls into the same projective family as $\calM^{1}$, so by property 2 of Lemma \ref{CMSZ}, step 3(c)(ii) will return $p'_0 < \xi-\eps_0-\eps$ with probability at most $\delta$.
        \item $\ell > 1$: If $\calH_m$ did not abort in the first $\ell-1$ rounds, then the last invocation of $\CheckCoins$ in the $(\ell-1)$st round returned some $p > \xi - \ell\eps_0$. Denote by $z_{\ell-1}$ the response in step 3(d) in the $(\ell-1)$st round. Fact \ref{fact:deferring_measurement} implies that the state at the beginning of the $\ell$th round looks identical to a state obtained after computing $\ValEst$ on some state and obtaining probability $p \geq \xi - \ell\eps_0$. Since this $\ValEst$ measurement falls into the same projective family as $\calM^{\ell}$, again by property 2 of Lemma \ref{CMSZ}, step 3(c)(ii) will return $p < \xi-\ell\eps_0-\eps$ with probability at most $\delta$.
    \end{itemize}

    Now, for iteration $s$, if step 3(c)(ii) returned some value $p \geq \xi-\ell\eps_0-(4(s-1)+1)\eps$, then by property 2 of Lemma \ref{lem:new-csmz}, the probability that step 3(c)(iii) aborts during iteration $s$ is at most $N_{\eps, \delta} \cdot (\eta + 4\delta)$, where $N_{\eps, \delta} = O\left(\frac{1}{\eps^2} \log \frac{1}{\delta}\right)$ is the number of possible outcomes of $\CheckCoins$ (see Lemma \ref{CMSZ}).
    
    Next, we let $x_s$ denote the probability that step 3(c)(iv) skips to step 3(d) in iteration $s$. By property 3 of Lemma \ref{lem:new-csmz} and a simple Markov argument,  
    \begin{align*}
    \Pr[\text{step 3(c)(ii) in iter. $s+1$ gives $p < \xi-\ell\eps_0-(4s+1)\eps$} \mid \text{not aborting in iter. $s$}] \leq \frac{N_{\eps, \delta}(\eta+4\delta)}{1-x_s}.
    \end{align*}
    But since we do not skip in iteration $s$ to step 3(d) with precisely probability $1-x_s$, conditioned on step 3(c)(ii) being good in iteration $s$, the probability that we fail in step 3(c)(iv) \emph{and} the next call to step 3(c)(ii) returns a bad probability is at most $N_{\eps, \delta} \cdot (\eta+4\delta)$.
    
    Iterating this argument, we see that for a given round $\ell$, the total probability that $\calH_m$ aborts in step 3(c)(iii) and step 3(c)(ii) returns a value below what we expect (unless step 3(c)(iv) succeeds, in which case we are done with round $\ell$) is at most $\delta + \iter \cdot 2N_{\eps, \delta}(\eta + 4\delta) \leq 1/km + \negl(\secp)$.
    
    Otherwise, we can assume that in iteration $s$, no abort has occurred in step 3(c)(iii) up to iteration $s$ in round $\ell$. In iteration $s$, given that the last $\Prepare$ returned a $p \geq \xi - \ell\eps_0 -(4s+1)\eps$, we have by property 2 of Lemma \ref{lem:new-csmz} that any subsequent $\ValEst$ would return a $p \geq \xi - \ell\eps_0 -(4s+2)\eps$ with probability at least $1-\delta$. But since such a $\ValEst$ is on expectation equal to the success probability of the prover's state at the start of step 3(c)(iv), this success probability is at least 
        \[ (\xi - \ell\eps_0 - (4s+2) \cdot \eps) \cdot (1-\delta) - \frac{1}{2} \cdot \delta \geq \xi - \ell\eps_0 - (4 \cdot \iter +2) \cdot \eps - \frac{3\delta}{2} \geq \xi-(\ell+1/2)\eps_0-\frac{3\delta}{2}. \]
    If we denote the prover's state after step 3(c)(iii) by $\bfrho$, it follows from property 1 of Lemma \ref{CMSZ} that
        \[ \bbE_{\bar{q} \gets Q_\ell^{(k)}}\left[p \hspace{3pt} \middle| \hspace{5pt} (p, \bfrho') \gets \ValEst_{V_{\tau_\ell}, U^{(\ell+1)}}(U_{\ell}(\bar{q}_\ell)\bfrho, \eps, \delta)\right] \geq \xi-(\ell+1/2)\eps_0-\frac{3\delta}{2}. \]
    Thus, by a simple Markov argument, it follows that step 3(c)(iv) will succeed with probability at least $\eps_0/3-\delta$ because $\bar{q}$ is sampled from the uniform distribution. Therefore, we can apply a Chernoff bound and bound the probability of aborting in step 3(c)(iv), which requires $\iter = \secp/\eps_0$ consecutive failures, by $\negl(\secp)$.

    Thus, by a union bound, the probability of aborting in step 3 is at most
    \[ m \cdot (1/km + \negl(\secp)) + m \cdot \negl(\secp) = 1/k+\negl(\secp), \]
    provided $m = \poly(\secp)$.
\end{proof}

Combining the two claims with another a union bound gives an overall probability of at most
    \[ \negl(\secp) + (1/k+\negl(\secp)) = 1/k+\negl(\secp), \]
as desired.
\end{proof}

Next, if we let $\calH_j$ also denote the distribution of outputs that a random execution of $\calH_j$ induces, then we claim the following is true:
\begin{lemma}\label{hybrids_public_sd}
    For $0 \leq j \leq m-1$,
    $\TD(\calH_j, \calH_{j+1}) \leq \sqrt{-\log(\xi/3m^2)/k} + 1/km + \negl(\secp)$.
\end{lemma}
Before we prove Lemma \ref{hybrids_public_sd}, we first show how Lemmas \ref{public_abort}, \ref{final_public_success}, and \ref{hybrids_public_sd} can be used to establish soundness.

Observe that in hybrid $\calH_m$, $\tau_m$ is \emph{independent} of $i$. Therefore, we have that 
\begin{align*}
    \Pr_{(i, \tau_m) \gets \calH_m}[\Accept_i(\tau_m) = 1] &\geq \Pr_{(i, \tau_m) \gets \calH_m}[\Accept_i(\tau_m) = 1 \mid \Accept(\tau_m) = 1] \cdot \Pr_{(i, \tau_m) \gets \calH_m}[\Accept(\tau_m) = 1] \\
    &= \frac{t}{k} \cdot \Pr_{(i, \tau_m) \gets \calH_m}[\Accept(\tau_m) = 1].
\end{align*}
Consequently, it follows that
\begin{align*}
    \Pr[\langle(\calA, \widetilde{V})(\secp, x)\rangle = 1] &= \Pr_{(i,\tau_m) \gets \calH_0}[\Accept_i(\tau_m) = 1] \\
    &\geq \Pr_{(i, \tau_m) \gets \calH_m}[\Accept_i(\tau_m) = 1] - \sum_{j=0}^{m-1} \TD(\calH_{j+1}, \calH_{j})\\
    &\geq \frac{t}{k} \cdot \Pr_{(i, \tau_m) \gets \calH_m}[\Accept(\tau_m) = 1]  - \sum_{j=0}^{m-1} \TD(\calH_{j+1}, \calH_{j}) \\
    &\geq \frac{t}{k}\left(1-\Pr_{(i,\tau_m) \gets \calH_m}[\tau_m = \bot]\right)  - \sum_{j=0}^{m-1} \TD(\calH_{j+1}, \calH_{j}) \\
    &\geq \frac{t}{k}-\frac{1}{k}-\negl(\secp)-m \cdot \left[\sqrt{\frac{-\log(\xi/3m^2)}{k}}+\frac{1}{km}+\negl(\secp)\right] \\
    &\geq \frac{t}{k}-\frac{1}{k}-\negl(\secp)-m \cdot \left[\sqrt{\frac{-\log(\xi/3m^2)}{k}}+\frac{1}{km}+\negl(\secp)\right] \\
    &\geq \frac{t}{k}-\frac{2}{k}-m\sqrt{\frac{-\log(\xi/3m^2)}{k}}-\negl(\secp), \\
    &\geq \frac{t}{k}-2m\sqrt{\frac{-\log(\xi/3m^2)}{k}}-\negl(\secp),
\end{align*}
as desired, which concludes the proof of Theorem \ref{public_coin_reduction}.
\end{proof}

Thus, it remains to prove Lemma \ref{hybrids_public_sd}:
\begin{proof}[Proof of Lemma \ref{hybrids_public_sd}]
    It is easy to see that in the first $j-1$ rounds, $\calH_j$ and $\calH_{j+1}$ are identical. Note that the sampling of queries in rounds $j+1$ and onwards is also identical, so it suffices to show that at the end of the $j$th round (i.e. when step 3(d) is reached), the prover states in $\calH_j$ and $\calH_{j+1}$ are close in trace distance.
    
    In the case of an abort in step 3(c)(iv) when $s = \iter$, we define the ending prover state to be any arbitrary quantum state (denoted by $\bot$). Without loss of generality, we can assume that both $\calH_j$ and $\calH_{j+1}$ start with the same fixed quantum state at the beginning of the $j$th iteration of step 3. 

    We now consider a new series of hybrids relating $\calH_j$ to $\calH_{j+1}$. For every $h\in[\iter]$, define $\calH_{j, h}$ to be the algorithm which behaves identically to $\calH_j$ except when $\ell = j$ and $s\in\{1,\ldots,h\}$, it replaces the conditional sampling of $\bar{q}$ in step 3(c)(i) with a uniform sample. By definition, we have that $\calH_{j, 0} := \calH_j$ and $\calH_{j, \iter} := \calH_{j+1}$. Now, for each hybrid $\calH_{j, h}$,  let $\calD_h$ denote the distribution of prover states at the end of the $j$th round and $\boldsymbol{s}_h$ denote the distribution of stopping times, which is the iteration on which the $\CheckCoins$ procedure first succeeds in round $j$ (or $\iter + 1$ if $\CheckCoins$ never succeeds).

    Our goal is to show that $\calH_{j, 0}$ and $\calH_{j, \iter}$ are close in trace distance:
    \begin{equation}\label{public_TD}
        \TD(\calH_{j, 0}, \calH_{j, \iter}) \leq \sqrt{-\log(\xi/3m^2)/k}+1/km+\negl(\secp).
    \end{equation}
    We prove Equation \eqref{public_TD} by showing that for every $h \in [\iter]$, $\calH_{j, h-1}$ and $\calH_{j, h}$ are close in trace distance:
    \begin{claim}\label{public_TD_hybrid}
        For all $1 \leq h \leq \iter$,
            \[ \TD(\calH_{j, h-1}, \calH_{j, h}) \leq \Pr[\boldsymbol{s}_{h-1} = h] \cdot \sqrt{\frac{-\log(\xi/3m^2)}{k}} + \frac{1}{km \cdot \iter} + \negl(\secp). \]
    \end{claim}
    \begin{proof}[Proof of Equation \eqref{public_TD} assuming Claim \ref{public_TD_hybrid}]
        We first observe that for all $h < h'$, $\calH_{j, h}$ and $\calH_{j, h'}$ behave identically in the $\ell$th round up until the $h$th iteration (when $s = h$). This means that the trace distance between the states in $\calH_{j, h}$ and $\calH_{j, h'}$ up until step 3(c)(iv) in iteration $h$ is zero. Additionally, since the conditionally sampled query in the $(h+1)$th iteration in $\calH_{j, h}$ is drawn from an identical distribution as the uniformly sampled query given an arbitrary starting quantum state (since it has no dependence on $i$), for all $s' \leq h+1$, $\Pr[\boldsymbol{s}_{h} = s'] = \Pr[\boldsymbol{s}_{h'} = s']$.

        Therefore, we have that
        \begin{align*}
            \TD(\calH_{j, 0}, \calH_{j, \iter}) &\leq \sum_{h=1}^{\iter} \TD(\calH_{j, h-1}, \calH_{j, h}) \\
            &\leq \sum_{h=1}^{\iter} \left(\Pr[\boldsymbol{s}_{h-1} = h] \cdot \sqrt{\frac{-\log(\xi/3m^2)}{k}} + \frac{1}{km \cdot \iter} + \negl(\secp)\right) \\
            &= \sum_{h=1}^{\iter} \left(\Pr[\boldsymbol{s}_{\iter} = h] \cdot \sqrt{\frac{-\log(\xi/3m^2)}{k}}\right) + \frac{1}{km} + \negl(\secp) \\
            &\leq \sqrt{\frac{-\log(\xi/3m^2)}{k}}+\frac{1}{km} + \negl(\secp),
        \end{align*}
        as desired.
    \end{proof}
    We now prove Claim \ref{public_TD_hybrid}:
    \begin{proof}[Proof of Claim \ref{public_TD_hybrid}]
        Fixing $h$, we observe (as noted earlier) that $\calH_{j, h-1}$ and $\calH_{j, h}$ behave identically in the first $h-1$ iterations, so $\TD(\calD_{h-1}|_{\boldsymbol{s}_{h-1} < h}, \calD_{h}|_{\boldsymbol{s}_{h} < h}) = 0$ and thus $\Pr[\boldsymbol{s}_{h-1} < h] = \Pr[\boldsymbol{s}_{h} < h]$. Additionally, since the first $h-1$ iterations have no dependence on $i$, we also have (as noted earlier) that $\Pr[\boldsymbol{s}_{h-1} = h] = \Pr[\boldsymbol{s}_{h} = h]$.

        Without loss of generality, we can fix the state $\bfrho$ in $\calH_{j, h-1}$ before the start of the $h$th iteration (and assume this is the same starting state as in $\calH_{j, h}$). A simple coupling argument now provides a bound on the desired trace distance:
        \begin{align*}
            \TD(\calH_{j, h-1}, \calH_{j, h}) &\leq \Pr[\boldsymbol{s}_{h-1} = h] \cdot \TD(\calH_{j, h-1}|_{\boldsymbol{s}_{h-1} = h}, \calH_{j, h}|_{\boldsymbol{s}_{h} = h}) \\
            &+ \Pr\left[ \boldsymbol{s}_{h-1} > h \right] \cdot \TD(\calH_{j, h-1}|_{\boldsymbol{s}_{h-1} > h}, \calH_{j, h}|_{\boldsymbol{s}_{h} > h)}).
        \end{align*}
        We now bound this expression with the following two propositions:
        \begin{proposition}\label{raz_public}
            $\TD(\calH_{j, h-1}|_{\boldsymbol{s}_{h-1} = h}, \calH_{j, h}|_{\boldsymbol{s}_{h} = h}) \leq \sqrt{-\log(\xi/3m^2)/k} + \negl(\secp)$.
        \end{proposition}
        \begin{proof}[Proof of Proposition \ref{raz_public}]
            First, it is easy to see that $\calH_{j, s'-1}$ and $\calH_{j, s'}$ behave identically before the $s'$th iteration. Thus, it suffices to bound the statistical distance between the joint distributions of accepting projections/queries with $i$, as the mixed states after continuing to the next round in both hybrids are entirely determined by the state before the beginning of the previous round and the queries being tested.\footnote{Note that although round $j+1$ of our hybrids use $\bar{q}_j$, upon acceptance, $\bar{q}_j$ itself contains no information about $i$, so applications of $\sfM^{j+1}$ in step 3(c) leak no additional information about $i$.} 
    
            Defining $W$ to be the event which occurs when the $\CheckCoins$ returns some $p$ such that $p \geq \xi - (\ell+1) \eps_0$ given a fixed state $\bfrho$ and projection $\Pi_{\bar{q}}$, we have that 
                \[ \TD(\calH_{j, h-1}|_{\boldsymbol{s}_{h-1} = h}, \calH_{j, h}|_{\boldsymbol{s}_{h} = h}) = \underset{i\gets [k]}{\bbE}[\TD((i, (Q_{\ell}^{(k)} | W)),(i, (Q_{\ell}^{(k)} | W, (Q_{\ell}^{(k)})_i)))]. \]
            Since $\Pr[W] \geq \eps_0/3-\negl(\secp)$ on average across all queries as previously argued by a simple application of a Markov bound, by Lemma \ref{raz_lemma} we have that
            \begin{align*}
                \TD(\calH_{j, h-1}|_{\boldsymbol{s}_{h-1} = h}, \calH_{j, h}|_{\boldsymbol{s}_{h} = h}) &= \underset{i\gets [k]}{\bbE}[\TD((i, (Q_{\ell}^{(k)} | W)),(i, (Q_{\ell}^{(k)} | W, (Q_{\ell}^{(k)})_i)))] \\
                &\leq \sqrt{-\log(\Pr[W])/k} \leq \sqrt{\log(3/\eps_0+\negl(\secp))/k} \\
                &= \sqrt{\frac{-\log(\xi/3m^2)}{k}}+\negl(\secp).
            \end{align*}
        \end{proof}
        \begin{proposition}\label{flooding_public}
            $\Pr[\boldsymbol{s}_{h-1} > h] \cdot \TD(\calH_{j, h-1}|_{\boldsymbol{s}_{h-1} > h}, \calH_{j, h}|_{\boldsymbol{s}_h > h}) \leq \frac{1}{km \cdot \iter}+\negl(\secp)$.
        \end{proposition}
        \begin{proof}[Proof of Proposition \ref{flooding_public}]
           When $h < \iter$, as noted earlier, it suffices to consider the state at the start of the $h$th iteration of step 3(c). Observing that this state has no dependence on $i$ or $q_j$, this implies that the sampling of $\Pi$ in both hybrids is drawn from the same uniform distribution over $\calP$. Therefore, we can apply Lemma \ref{lem:new-csmz} and note that by a coupling argument, we have that 
               \[ \TD(\calH_{j, h-1}|_{\boldsymbol{s}_{h-1} > h}, \calH_{j, h}|_{\boldsymbol{s}_h > h}) \leq \TD(\calD_{h-1}|_{\boldsymbol{s}_{h-1} > h}, \calD_{h}|_{\boldsymbol{s}_h > h}) \leq \frac{\TD(\calD_{h-1}, \calD_{h})}{\Pr[\boldsymbol{s}_{h-1} > h]} \leq \frac{N_{\eps, \delta} \cdot \eta}{\Pr[\boldsymbol{s}_{h-1} > h]}, \]
           and so 
               \[ \Pr[\boldsymbol{s}_{h-1} > h] \cdot \TD(\calH_{j, h-1}|_{\boldsymbol{s}_{h-1} > h}, \calH_{j, h}|_{\boldsymbol{s}_h > h}) \leq N_{\eps, \delta} \cdot \eta \leq \frac{1}{km \cdot \iter}. \]
           On the other hand, through an identical argument as in Lemma \ref{public_abort}, we observe that
               \[ \Pr[\boldsymbol{s}_{\iter-1} > \iter] = \negl(\secp), \]
           and thus
               \[ \Pr[\boldsymbol{s}_{\iter-1} > \iter] \cdot \TD(\calH_{j, \iter-1}|_{\boldsymbol{s}_{\iter-1} > \iter}, \calH_{j, \iter}|_{\boldsymbol{s}_\iter > \iter}) \leq \Pr[\boldsymbol{s}_{\iter-1} > \iter] \cdot 1 = \negl(\secp), \]
           which finishes the proof.
        \end{proof}
        
        Thus, we have that 
        \begin{align*}
            \TD(\calH_{j, h-1}, \calH_{j, h}) &\leq  \Pr[\boldsymbol{s}_{h-1} = h] \cdot \TD(\calH_{j, h-1}|_{\boldsymbol{s}_{h-1} = h}, \calH_{j, h}|_{\boldsymbol{s}_{h} = h}) \\
            &+ \Pr\left[\boldsymbol{s}_{h-1} > h\right] \cdot \TD(\calH_{j, h-1}|_{\boldsymbol{s}_{h-1} > h}, \calH_{j, h}|_{\boldsymbol{s}_{h} > h}) \\
            &\leq \Pr[\boldsymbol{s}_{h-1} = h] \cdot \sqrt{\frac{-\log(\xi/3m^2)}{k}} + \frac{1}{km \cdot \iter} + \negl(\secp),
        \end{align*}
        as claimed, concluding the proof of Claim \ref{public_TD_hybrid}.
    \end{proof}
    Since Equation \eqref{public_TD} is equivalent to our claim, this concludes the proof of Proposition \ref{hybrids_public_sd}.
\end{proof}

\section{Parallel Repetition for Three-Message Protocols}\label{sec:three-message}
In this section, we analyze the soundness of parallel repetition of private-coin protocols that consist of three messages. Although exponentially-decreasing soundness from parallel repetition for three-message arguments was already demonstrated through other means in \cite{BQSY24}, we provide a simpler proof of parallel repetition (for protocols where the verifier is classical) and generalize from the direct-product setting to the setting of threshold verifiers.

In the private-coin setting, one cannot compute the predicate of the external verifier, and so a proxy for the external verifier's verdict must be used. Existing classical reductions take two general approaches: the correlation reduction strategy of \cite{CHS05} and the \emph{soft-decision} procedure of \cite{BIN97} and \cite{HPPW10}. \cite{BQSY24} provides a quantum analogue of correlation reduction, while we consider an analogue of the soft-decision procedure (see Section \ref{sec:overview} for more details).

\subsection{Our Results}
We prove the following parallel repetition theorem for three-message protocols:
\begin{theorem}\label{three_message_reduction}
    Let $(P, V)$ be a three-message interactive argument. Suppose that there exists a quantum adversary $\calB$, $\xi=\xi(\secp) > 0$, and polynomials $1 \leq t = t(\secp) \leq k=k(\secp)$ such that for all $\secp \in \mathbb{N}$,
    \begin{equation}\label{three_message_parallel_cheater}
        \Pr[\langle (\calB, V^{(t, k)})(1^\secp) \rangle = 1] \geq \xi.
    \end{equation}
    Then there exists a quantum adversary $\calA$ running in time $\poly(|\calB|, \secp, 1/\xi)$ and a negligible function $\negl$ such that for every $\secp\in \mathbb{N}$,
    \[
        \Pr[\langle (\calA, V)(1^\secp) \rangle = 1] \geq \frac{t}{k}-\frac{2\log k}{\sqrt{k}}-3\sqrt{\frac{-\log\xi}{k}}-\negl(\secp).\footnote{We remark that the precise constants here can be improved by tweaking the parameters of our reduction, but we will keep these for ease of presentation.}
    \]
\end{theorem}
Note that Theorem \ref{three_message_reduction} implies the following corollary:
\begin{corollary}\label{three_message_parallel_repetition} 
    Let $(P, V)$ be a three-message interactive argument and let $\eps = \eps(\secp)$ be any parameter such that for every $\QPT$ adversary $\calA$ there exists a negligible function $\negl$ such that for every $\secp \in \mathbb{N}$,
    \[
        \Pr[\langle (\calA, V)(1^\secp) \rangle = 1] \leq \eps(\secp)+\negl(\secp).
    \]
    Fix any polynomials $1 \leq t = t(\secp) \leq k = k(\secp)$ such that $\eps < \frac{t}{k}-\frac{2\log k}{\sqrt{k}}$.\footnote{The condition that $\eps < \frac{t}{k}-\frac{2\log k}{\sqrt{k}}$ is similar to the condition in Corollary \ref{public_coin_parallel_repetition} but with some minor loss.} Then for every $\QPT$ adversary $\calB$ there exists a negligible function $\negl$ such that for every $\secp\in\mathbb{N}$,
    \[
        \Pr[\langle (\calB, V^{(t, k)})(1^\secp) \rangle = 1] \leq \max\{2\exp\left(\frac{-k}{9} \cdot \left(\frac{t-2\sqrt{k}\log k}{k}-\eps\right)^2\right), \negl(\secp)\}.
    \]
\end{corollary}
\begin{proof}[Proof of Corollary \ref{three_message_parallel_repetition}]
    Suppose for the sake of contradiction that there exists a $\QPT$ adversary $\calB$ such that $\calB$ succeeds against $V^{(t, k)}$ with non-negligible probability $\xi > 2\exp\left(\frac{-k}{9} \cdot \left(\frac{t-2\sqrt{k}\log k}{k}-\eps\right)^2\right)$. Applying Theorem \ref{three_message_reduction} then gives a quantum adversary $\calA$ running in time $\poly(\secp, 1/\xi)$ such that
    \begin{align*}
        \Pr[\langle (\calA, V)(1^\secp) \rangle = 1] &\geq \frac{t}{k}-\frac{2\log k}{\sqrt{k}}-3\sqrt{\frac{-\log(2\exp\left(-\frac{k}{9} \left(\frac{t}{k}-\frac{2 \log k}{\sqrt{k}}-\eps\right)^2\right))}{k}}-\negl(\secp) \\
        &= \frac{t}{k}-\frac{2\log k}{\sqrt{k}}-3\sqrt{\frac{-\log(\exp\left(-\frac{k}{9} \left(\frac{t}{k}-\frac{2\log k}{\sqrt{k}}-\eps\right)^2\right))-1}{k}}-\negl(\secp) \\
        &= \frac{t}{k}-\frac{2\log k}{\sqrt{k}}-3\sqrt{\frac{\frac{k}{9} \cdot \left(\frac{t}{k}-\frac{2\log k}{\sqrt{k}}-\eps\right)^2-1}{k}}-\negl(\secp) \\
        &= \frac{t}{k}-\frac{2\log k}{\sqrt{k}}-\sqrt{\left(\frac{t}{k}-\frac{2\log k}{\sqrt{k}}-\eps\right)^2-\frac{9}{k}}-\negl(\secp).
    \end{align*}
    Observe that since $\xi \leq 1$, this means that $c := 2\exp\left(\frac{-k}{9} \cdot \left(\frac{t-2\sqrt{k}\log k}{k}-\eps\right)^2\right) < \xi \leq 1$ and so $3\sqrt{\frac{-\log c}{k}} = \sqrt{\left(\frac{t}{k}-\frac{2\log k}{\sqrt{k}}-\eps\right)^2-\frac{9}{k}} > 0$. Since $\sqrt{x^2-y} \leq x-\frac{y}{2x}$ and hence $-\sqrt{x^2-y} \geq -x+\frac{y}{2x}$ for $x, y > 0$ such that $x^2-y > 0$, we have that
    \begin{align*}
        \Pr[\langle (\calA, V)(1^\secp) \rangle = 1] &\geq \frac{t}{k}-\frac{2\log k}{\sqrt{k}}-\sqrt{\left(\frac{t}{k}-\frac{2\log k}{\sqrt{k}}-\eps\right)^2-\frac{9}{k}}-\negl(\secp) \\
        &\geq \frac{t}{k}-\frac{2\log k}{\sqrt{k}}-\left(\frac{t}{k}-\frac{2\log k}{\sqrt{k}}-\eps\right)+\frac{9/k}{2\left(\frac{t}{k}-\frac{2\log k}{\sqrt{k}}-\eps\right)}-\negl(\secp) \\
        &= \eps+\frac{9/k}{2\left(\frac{t}{k}-\frac{2\log k}{\sqrt{k}}-\eps\right)}-\negl(\secp) \\
        &\geq \eps+\frac{9/k}{2t/k}-\negl(\secp) = \eps+\frac{9}{2t}-\negl(\secp).
    \end{align*}
    Since $t$ is polynomial, $\eps+\frac{9}{2t}-\negl(\secp) > \eps+\negl(\secp)$.
    
    But if $\calB$ has non-negligible advantage, then $1/\xi$ is polynomial in $\secp$ and so $\calA$ is $\QPT$, in contradiction to the original assumption that no such $\QPT$ adversary exists. Thus, no such $\QPT$ adversary $\calB$ exists either.
\end{proof}

\begin{proof}[Proof of Theorem \ref{three_message_reduction}]
    Denote by $\ket{\psi}$ the initial state of $\calB$, and let $\Lambda = \poly(\secp)$ be the number of qubits in $\ket{\psi}$. For the sake of simplicity, we assume without loss of generality that $\ket{\psi}$ contains designated registers $\calZ_1$ and $\calZ_2$, where $\calZ_j$ is the register on which $\calB$ writes its $j$th message. Since the prover's unitary in the first round is efficiently implementable and independent of the verifier's queries, we also assume without loss of generality that $\calB$ determines its first round message by simply measuring the $\calZ_1$ register of $\ket{\psi}$. We denote all the other registers in $\ket{\psi}$ by $\calI$.
    
    We denote by $Q$ the distribution of queries of the single-fold verifier in the protocol $(P, V)$; let $Q^{(k)}$ refer to the distribution of the queries sent in the $k$-fold repeated protocol $(P^{(k)},V^{(t, k)})$ and $\bar{q}^i$ refer to the $i$th coordinate of the query vector $\bar{q}$.

    We denote by $U_{\bar{q}}$ (which operates on registers $\calI$ and $\calZ_2$) the unitary that $\calB$ applies to its quantum state upon receiving query $\bar{q}$ before measuring the $\calZ_2$ register to compute its second round response. As in Section \ref{sec:public-coin}, the assumption that $U_{\bar{q}}$ is a unitary is without loss of generality, as any adversary not of this form can be ``purified'' into an adversary with unitary strategies, identical observable behavior, and constant factor blowup in size. 

    For a $k$-fold transcript $\tau = (\bar{z}_1, \bar{q}, \bar{z}_2)$, where $\bar{q}$ is generated with verifier internal randomness $\bar{r} = (\bar{r}^1, \ldots, \bar{r}^k)$, let $\tau^j = (\bar{z}_1^j, \bar{q}^j, \bar{z}_2^j)$ and $\bar{r}^j$ be the view and internal randomness of the $j$'th execution. Let $\Accept_j(\bar{r}, \tau) = 1$ if and only if the (single-execution) verifier accepts $\tau^j$ with respect to internal randomness $\bar{r}^j$. Similarly, given the verifier's internal randomness $\bar{r}^{(-i)}$ in all but the $i$th execution, one can define $\Accept_j(\bar{r}^{(-i)}, \tau) = 1$ for $j \neq i$ to occur exactly when the (single-execution) verifier accepts $\tau^j$ with respect to internal randomness $\bar{r}^j$. Again, for fixed threshold $t$ we define $\Accept(\bar{r}, \tau) = 1$ iff $\sum_i \Accept_i(\bar{r}, \tau) \geq t$. We also define the classical function $\SoftDecision_{\nu, t}(i, \bar{r}^{(-i)}, \tau, \omega)$ parametrized by a threshold $0 \leq t \leq k$ and smoothness parameter $\nu > 0$ which takes as input an index $i \in [k]$, verifier internal randomness $\bar{r}^{(-i)}$, transcript $\tau$, and random string $\omega \in \{0, 1\}^{\poly(\secp)}$, and outputs a bit $b \in \{0, 1\}$ such that if $\sum_{j \neq i} \Accept_j(\bar{r}^{(-i)}, \tau) = \ell$, then 
    \[ 
        \Pr_{\omega \gets \{0, 1\}^{\poly(\secp)}}[\SoftDecision_{\nu, t}(i, \bar{r}^{(-i)}, \tau, \omega) = 1] = \min\{1, 2^{\nu(\ell+1-t)}\}. 
    \]
    One can think of $\SoftDecision$ as using $\omega$ as its random coins to decide whether to accept with probability $\min\{1, 2^{\nu(\ell+1-t)}\}$.

    \begin{remark}\label{remark:computing_private_queries}
        Recall that in Lemma \ref{CMSZ}, $\ValEst$ is defined with respect to a verdict function $V$ and adversary circuit $A$, where $A$ takes as input randomness $u$ and returns a response $z$, and $V$ is a deterministic function of $u$ and $z$. For our setting, we take $A := A_{\bar{z}_1}$ to be the algorithm that has the first message $\bar{z}_1$ of $\calB$ hardwired and on input $\bar{r}$, which represents the verifier's random coins, computes the verifier's message $\bar{q}$ corresponding to the randomness $\bar{r}$, and then applies $U_{\bar{q}}$ to obtain $\bar{z}_2$, and outputs $(\bar{q}, \bar{z}_2)$. Now, we define $V := V_{\bar{z}_1}$ to be the verdict function which takes as input $\bar{r}, (\bar{q}, \bar{z}_2)$ and returns $\Accept(\bar{r}, \bar{z}_1||\bar{q}||\bar{z}_2)$.\footnote{Note that $\bar{q}$ is technically redundant since it can be computed from $\bar{r}$ and $\bar{z}_1$. We include it only for notational convenience.} Note that the estimated success probability in this $\ValEst$ is indeed the same success probability of the original prover $\calB$ in the three-message protocol. 
    \end{remark}
    
    The general framework of our reduction is the same as in the public-coin setting, so we refer the reader to the informal explanation given in Section \ref{sec:public-coin} for some intuition behind the reduction in the three-message setting as well.
    
    Taking $\ValEst$, $\Prepare$, and $\Repair'$ to be the algorithms defined in Lemmas \ref{CMSZ} and \ref{lem:new-csmz}, our algorithm $\calA$ is described in the figure below (Figure \ref{alg:A-three-desc}).
    \defbox{Projection}{$\SoftDecisionProj_{\bar{z}_1, j, \bar{r}^{(-j)}, q, \omega}[\bfrho]$}{SoftDecision}{
        \begin{enumerate}
            \item
                Use $\bar{r}^{(-j)}$ to compute query $\bar{q}^{(-j)}$ and complete $\bar{q}^{(-j)}$ with $\bar{q}^j = q$ to get query $\bar{q}$.
            \item 
                Apply $U_{\bar{q}}$ to $\bfrho$, resulting in state $\bfrho' = \sum_{\bar{z}_2} (\bfrho'_{\bar{z}_2})_{\calI, \calZ_1} \otimes \ket{\bar{z}_2}\bra{\bar{z}_2}_{\calZ_2}$. 
            \item
                Apply the classical function $\SoftDecision_{\nu, t}(j, \bar{r}^{(-j)}, \bar{z}_1 || \bar{q} || \cdot, \omega)$ to $\brho'\otimes\ket{0}\bra{0}_\calE$ coherently over responses $\bar{z}_2$ and write the outcome in register $\calE$, resulting in state $\sum_{\bar{z}_2} (\bfrho'_{\bar{z}_2})_{\calI, \calZ_1} \otimes \ket{\bar{z}_2}\bra{\bar{z}_2}_{\calZ_2} \otimes \ket{b_{\bar{z}_2}}\bra{b_{\bar{z}_2}}_{\calE}$.
            \item
                Measure and discard register $\calE$, obtaining $b$ and remaining state $\bfrho''$.
            \item
                Apply $U_{\bar{q}}^{\dagger}$ to get state $\bsigma = \bfrho'''_{\calI, \calZ_1, \calZ_2}$ before returning $(\bfrho''', b)$.
        \end{enumerate}
    }
    \algo{The Algorithm $\calA$}{A-three-desc}{    
        Let $\iter := 4\secp/\xi$ and $\Lambda := \mathsf{size}(\ket{\psi})$ be the number of qubits of $\ket{\psi}$. $\calA$ begins with state $\ket{\psi}^{\otimes \iter}$, and runs as follows:
        
        \begin{enumerate}
        \item
            Set $\eps_0 := \xi/4, \eps := \frac{\eps_0}{16 \cdot \iter}, \delta := \min\{2^{-\secp}, 2^{-k}\}, \eta := \frac{1}{4k \cdot \iter}/N_{\eps, \delta}$, $\nu := \sqrt{-\log{\xi}/k}$, where $N_{\eps, \delta}$ is the parameter from Lemma \ref{CMSZ}. Sample $i \gets [k]$.
        \item
            For $s = 1$ to $\iter$:
            \begin{enumerate}
                \item[(a)]
                    Set $\bfrho = \ket{\psi}$ to be $\calA$'s $s$th copy of $\ket{\psi}$. Measure the $\calZ_1$ register to get response $\bar{z}_1$ and leftover state $\bfrho_0$. Run $(p_0, \bfrho_1) \gets \ValEst_{V_{\bar{z}_1}, A_{\bar{z}_1}}(\bfrho_0, \eps, \delta)$.
                    \begin{enumerate}
                        \item
                            If $p_0 \geq \xi - \eps_0$, fix $\bar{z}_1$ and send $\bar{z}_1^i$ to $V$ before continuing to step 3.
                        \item
                            Else, if $p_0 < \xi - \eps_0$ and $s = \iter$, return $(i, \bot)$.
                    \end{enumerate}
            \end{enumerate}
        \item
            Let $\calP := \{\SoftDecisionProj_{\bar{z}_1, j, \bar{r}^{(-j)}, q, \omega}\}_{j \in [k], \bar{r}^{(-j)} \in \{0, 1\}^{\poly(\secp)}, q \in Q, \omega \in \{0, 1\}^{\poly(\secp)}}$ be the projection family defined above. Define the almost projective measurement family $\sfM := \ValEst_{V_{\bar{z}_1}, A_{\bar{z}_1}}$. 
        \item 
            Receive the message $q$ from $V$. 
        \item For $s = 1$ to $\iter$: 
        \begin{enumerate}
            \item[(a)] Generate verifier randomness $\bar{r}^{(-i)} \gets \{0, 1\}^{\poly(\secp)}$ for executions $[k] \setminus \{i\}$ and generate a uniformly random string $\omega \gets \{0, 1\}^{\poly(\secp)}$. Let $\Pi := \SoftDecisionProj_{\bar{z}_1, i, \bar{r}^{(-i)}, q, \omega}$.
            \item[(b)] Compute $(\bfrho_1', p_1) \gets \sfM_{\eps, \delta}(\bfrho_1)$.
            \item[(c)] Compute $(\bfrho_1'', p'_1) \gets \Prepare_{\sfM, \calP}(\bfrho'_1, p_1, \eps, \delta, \eta, \Lambda)$. 
            
            If $p'_1 < \xi - \eps_0 - (4s+1) \eps$, return $(i, \bot)$.
            \item[(d)] 
            Compute $(\bsigma, b) \gets \Pi[\bfrho''_1]$.
            
            If $b = 1$, skip to step 6. 
            
            Else, if $b = 0$ and $s = \iter$, return $(i, \bot)$.
            \item[(e)] Compute $\bfrho_1 = \Repair'_{\sfM, \calP, \Pi}(\bsigma, b, p'_1, \eps, \delta, \eta, \Lambda)$.
        \end{enumerate}
        \item
            Use $\bar{r}^{(-i)}$ to compute query $\bar{q}^{(-i)}$ and complete $\bar{q}^{(-i)}$ with $\bar{q}^i = q$ to get query $\bar{q}$. Apply $U_{\bar{q}}$ to $\bsigma$ before measuring the $\calZ_2$ register to get response $\bar{z}_2$.
        \item
           Set $\tau = \bar{z}_1 || \bar{q}^{*} || \bar{z}_2$ and send $\bar{z}^i_2$ back to $V$.
        \item 
           Return $(i, \tau)$.
        \end{enumerate}
    }
\noindent
\textbf{Runtime.}
The fact that $\calA$ runs in time $\poly(|\calB|, \secp, 1/\xi)$ follows from the efficiency of $\ValEst'$, $\Prepare$, and $\Repair'$, and the fact that all parameters are polynomial in $\secp$ and $1/\xi$.
$ $\newline
\textbf{Soundness.}
Before we analyze the soundness of our reduction, we note that we can assume without loss of generality that $\xi \gg \delta$, as otherwise our guarantee is essentially trivially true. 

\noindent We begin by considering the following hybrids $\calH_0$ and $\calH_1$:
\begin{itemize}
    \item Hybrid $\calH_0$: Run a random execution of $\calA$ and return $(i, \tau)$.
    \item Hybrid $\calH_1$: In step 5(a), replace $i$ with a uniformly random index $j \gets [k]$, sample a fresh query $q \gets Q$, and define $\Pi := \SoftDecisionProj_{\bar{z}_1, j, \bar{r}^{(-j)}, q, \omega}$ instead. Alternatively, one can think of replacing the conditionally sampled projection $\Pi$ in step 5(a) with an unconditionally sampled projection $\Pi \gets \calP$. Having done this for all iterations, derive a full transcript $\tau$. Now, sample $i \gets [k]$ and return $(i, \tau)$.
\end{itemize}
Note that by definition, $\calH_0$ outputs a random transcript as produced by $(\calA, V)$, while $\calH_1$ outputs a transcript that has \emph{no dependence} on $i$.

It is easy to see that step 6 of $\calA$ undoes the effect of step 5 in $\SoftDecisionProj$, and since step 6 is only reached when an outcome of $b = 1$ is measured, this means that the state after measuring $\bar{z}_2$ has the property that $\SoftDecision_{\nu, t}(j, \bar{r}^{(-j)}, \bar{z}_1||\bar{q}||\bar{z}_2, \omega) = 1$.

We will mirror the argument from the proof of Theorem \ref{public_coin_reduction}, although with a slightly different ordering.

\begin{lemma}\label{final_three_success}
    $\underset{(i, \tau) \gets \calH_1}{\Pr} [\tau = \bot] \leq 1/2k+\negl(\secp)$.
\end{lemma}
\begin{proof}[Proof of Lemma \ref{final_three_success}]
We first bound the probability that $\calH_1$ aborts in step 2:
\begin{claim}\label{three_step_two}
    $\calH_1$ aborts in step 2 with probability at most $\negl(\secp)$.
\end{claim}
\begin{proof}[Proof of Claim \ref{three_step_two}]
    First, note that because the starting state $\ket{\psi}$ has success probability which is at least $\xi$, by property 1 of Lemma \ref{CMSZ} and Fact \ref{fact:deferring_measurement}, we have that 
        \[\bbE[p_0 | (p_0, \bfrho') \gets \ValEst_{V_{\bar{z}_1}, A_{\bar{z}_1}}(\bfrho_0, \eps, \delta)] \geq \xi.\]
    Recall that since $\ValEst$ takes on values in $[-\frac{1}{2}, \frac{3}{2}]$ (see Lemma \ref{CMSZ}), by a simple Markov argument, the probability that any particular iteration of step 2(a) will succeed is
    \begin{align*}
        \Pr\left[p_0 \geq \xi - \eps_0 \hspace{3pt} \middle| \hspace{5pt} (p_0, \bfrho') \gets \ValEst_{V_{\bar{z}_1}, A_{\bar{z}_1}}(\bfrho_0, \eps, \delta)\right] \geq 2\eps_0/3.
    \end{align*}
    Thus, by using a Chernoff bound we conclude that the probability of aborting in step 2 is at most $\negl(\secp)$.
\end{proof}
We now bound the probability of $\calH_1$ aborting in step 5(c) or 5(d):
\begin{claim}\label{three_step_five}
    Conditioned on not aborting in step 2, $\calH_1$ aborts in step 5 with probability at most $\frac{1}{2k}+\negl(\secp)$.
\end{claim}
\begin{proof}[Proof of Claim \ref{three_step_five}]
    We begin by analyzing step 5(b) when $s = 1$: since the last $\ValEst$ measurement in step 2(a) before $\calA$ moved on to step 3 falls into the same projective family as $\calM$, by property 2 of Lemma \ref{CMSZ}, step 5(b) will return $p_1 < \xi-\eps_0-\eps$ with probability at most $\delta$.

    Now, for all iterations $s \geq 1$, if step 5(b) returned some value $p_1 \geq \xi-\eps_0-(4(s-1)+1)\eps$, then by property 2 of Lemma \ref{lem:new-csmz}, the probability that step 5(c) aborts during iteration $s$ is at most $N_{\eps, \delta} \cdot (\eta + 4\delta)$, where $N_{\eps, \delta} = O\left(\frac{1}{\eps^2} \log \frac{1}{\delta}\right)$ is the number of possible outcomes of $\CheckCoins$ (see Lemma \ref{CMSZ}).

    Next, we let $x_s$ denote the probability that step 5(d) skips to step 6 in iteration $s$. By property 3 of Lemma \ref{lem:new-csmz} and a simple Markov argument,  
    \begin{align*}
        \Pr[\text{step 5(b) in iter. $s+1$ gives $p < \xi-\eps_0-(4s+1)\eps$} \mid \text{not aborting in iter. $s$}] \leq \frac{N_{\eps, \delta}(\eta+4\delta)}{1-x_s}.
    \end{align*}
    But since we do not skip in iteration $s$ to step 6 with precisely probability $1-x_s$, conditioned on step 5(b) being good in iteration $s$, the probability that we fail in step 5(d) \emph{and} step 5(b) is bad in iteration $s+1$ is at most $N_{\eps, \delta} \cdot (\eta+4\delta)$.

    Iterating this argument, we see that the total probability that $\calH_1$ aborts in step 5(c) and step 5(b) returns a value below what we expect in any iteration (unless step 5(d) succeeds, in which case we are done) is at most $\delta + \iter \cdot 2N_{\eps, \delta}(\eta + 4\delta) \leq 1/2k + \negl(\secp)$.

    Otherwise, we can assume that in iteration $s$, no abort has occurred in step 5(c) up to iteration $s$ in round $\ell$. In iteration $s$, given that the last $\Prepare$ returned a $p \geq \xi - \eps_0 - (4s+1)\eps$, we have by property 2 of Lemma \ref{lem:new-csmz} that any subsequent $\ValEst$ would return a $p \geq \xi - \eps_0 -(4s+2)\eps$ with probability at least $1-\delta$. But since such a $\ValEst$ is on expectation equal to the success probability of the prover's state at the start of step 5(d), this success probability is at least 
        \[ (\xi - \eps_0 - (4s+2) \cdot \eps) \cdot (1-\delta) - \frac{1}{2} \cdot \delta \geq \xi - \eps_0 - (4 \cdot \iter +2) \cdot \eps - \frac{3\delta}{2} \geq \xi-\frac{3\eps_0}{2}-\frac{3\delta}{2}. \]
    We now note that the event that $\sum_{j \neq i} \Accept_j(\bar{r}^{(-j)}, \tau) \geq t-1$ immediately implies that $\SoftDecision$ returns 1 (and in particular this also means that if $\sum_{j} \Accept_j(\bar{r}, \tau) \geq t$ then $\SoftDecision$ will return 1), so the probability of success at step 5(d) is at least the true success probability of the state at the start of step 5(d).
    Thus, the probability of succeeding at step 5(d) is always at least 
        \[ \xi-\frac{3\eps_0}{2}-\frac{3\delta}{2} \geq 2\eps_0, \]
    for sufficiently large $\secp$.
    Therefore, we can apply a Chernoff bound and bound the probability of aborting in step 5(d), which requires $\iter = \secp/\eps_0$ consecutive failures, by $\negl(\secp)$.

    The claim then follows immediately from a union bound.
\end{proof}

    The two claims, combined with a second union bound, bound the probability of $\calH_1$ aborting in any step by at most
        \[ \negl(\secp) +(1/2k + \negl(\secp)) = 1/2k+\negl(\secp), \]
    as desired.
\end{proof}

Next, if we let $\calH_j$ also denote the distribution of outputs that a random execution of $\calH_j$ induces, then we claim the following is true:
\begin{lemma}\label{hybrids_three_sd}
    $\TD(\calH_0, \calH_1) \leq \sqrt{-\log \xi/k} + 1/4k + \negl(\secp)$.
\end{lemma}
\begin{proof}[Proof of Lemma \ref{hybrids_three_sd}]
It is easy to see that step 1 of both $\calH_0$ and $\calH_1$ are identical, so we can assume without loss of generality that both $\calH_0$ and $\calH_1$ start with the same fixed quantum state at the beginning of step 5.

In the case of an abort in step 5(d) when $s = \iter$, we define the ending prover state to be any arbitrary quantum state (denoted by $\bot$). It suffices to show that if and when step 6 is reached, the prover states in $\calH_0$ and $\calH_1$ are close in trace distance.

We now consider a new series of hybrids relating $\calH_0$ to $\calH_1$. Define $\calH_{0, h}$ to be the algorithm which behaves identically to $\calH_0$ except when $1 \leq s \leq h$, it samples fresh queries $q \gets Q$ in step 5(a) when determining $\Pi$ instead of using the verifier's query. By definition, we have that $\calH_{0, 0} := \calH_0$ and $\calH_{0, \iter} := \calH_1$. Now, for each hybrid $\calH_{0, h}$, let $\calD_h$ denote the distribution of prover states at the end of step 6 (or steps 5(c)/5(d) if an abort has occurred) and $\boldsymbol{s}_h$ denote the distribution of stopping times, which is the iteration on which the $\SoftDecisionProj$ procedure first succeeds (or $\iter + 1$ if $\SoftDecisionProj$ never succeeds).

Therefore, our goal is to show that $\calH_{0, 0}$ and $\calH_{0, \iter}$ are close in trace distance:
\begin{equation}\label{three_TD}
    \TD(\calH_{0, 0}, \calH_{0, \iter}) \leq \sqrt{-\log \xi/k}+1/4k+\negl(\secp).
\end{equation}
We prove Equation \eqref{three_TD} by proving that for every $h \in [\iter]$, $\calH_{0, h-1}$ and $\calH_{0, h}$ are close in trace distance:
\begin{claim}\label{three_TD_hybrid}
    For all $1 \leq h \leq \iter$,
        \[ \TD(\calH_{0, h-1}, \calH_{0, h}) \leq \Pr[\boldsymbol{s}_{h-1} = h] \cdot \sqrt{-\log \xi/k} + \frac{1}{4k \cdot \iter} + \negl(\secp). \]
\end{claim}
\begin{proof}[Proof of Equation \ref{three_TD} assuming Claim \ref{three_TD_hybrid}]
    We first observe that for all $h < h'$, $\calH_{0, h}$ and $\calH_{0, h'}$ behave identically up until the $h$th iteration (when $s = h$). This means that the trace distance between the states in $\calH_{0, h}$ and $\calH_{0, h'}$ up until step 5(d) in iteration $h$ is zero. Additionally, since the projection $\Pi$ in the $(h+1)$th iteration in $\calH_{0, h}$ is drawn from the same distribution as the in the $(h+1)$th iteration of $\calH_{0, h'}$ given an arbitrary starting quantum state which has no dependence on $i$, for all $s' \leq h+1$, $\Pr[\boldsymbol{s}_{h} = s'] = \Pr[\boldsymbol{s}_{h'} = s']$.

    Therefore, we have that
    \begin{align*}
        \TD(\calH_{0, 0}, \calH_{0, \iter}) &\leq \sum_{h=1}^{\iter} \TD(\calH_{0, h-1}, \calH_{0, h}) \\
        &\leq \sum_{h=1}^{\iter} \left(\Pr[\boldsymbol{s}_{h-1} = h] \cdot \sqrt{-\log \xi/k} + \frac{1}{4k \cdot \iter} + \negl(\secp)\right) \\
        &= \sum_{h=1}^{\iter} \left(\Pr[\boldsymbol{s}_{\iter} = h] \cdot \sqrt{-\log \xi/k}\right) + \frac{1}{4k} + \negl(\secp) \\
        &\leq \sqrt{-\log \xi/k}+\frac{1}{4k} + \negl(\secp),
    \end{align*}
    as desired.
\end{proof}

We now prove Claim \ref{three_TD_hybrid}:
    \begin{proof}[Proof of Claim \ref{three_TD_hybrid}]
        Fixing $h$, we observe (as noted earlier) that $\calH_{0, h-1}$ and $\calH_{0, h}$ behave identically in the first $h-1$ iterations, so $\TD(\calD_{h-1}|_{\boldsymbol{s}_{h-1} < h}, \calD_{h}|_{\boldsymbol{s}_{h} < h}) = 0$ and thus $\Pr[\boldsymbol{s}_{h-1} < h] = \Pr[\boldsymbol{s}_{h} < h]$. Additionally, since the first $h-1$ iterations have no dependence on $i$, we also have (as noted earlier) that $\Pr[\boldsymbol{s}_{h-1} = h] = \Pr[\boldsymbol{s}_{h} = h]$.

        Without loss of generality, we can fix the state $\bfrho$ in $\calH_{0, h-1}$ before the start of the $h$th iteration (and assume this is the same starting state as in $\calH_{0, h}$). A simple coupling argument now provides a bound on the desired trace distance:
        \begin{align*}
            \TD(\calH_{0, h-1}, \calH_{0, h}) &\leq \Pr[\boldsymbol{s}_{h-1} = h] \cdot \TD(\calH_{0, h-1}|_{\boldsymbol{s}_{h-1} = h}, \calH_{0, h}|_{\boldsymbol{s}_{h} = h}) \\
            &+ \Pr\left[ \boldsymbol{s}_{h-1} > h \right] \cdot \TD(\calH_{0, h-1}|_{\boldsymbol{s}_{h-1} > h}, \calH_{0, h}|_{\boldsymbol{s}_{h} > h)}).
        \end{align*}
        We can now bound this expression with the following two propositions:
        \begin{proposition}\label{flooding_three}
            $\Pr[\boldsymbol{s}_{h-1} > h] \cdot \TD(\calH_{0, h-1}|_{\boldsymbol{s}_{h-1} > h}, \calH_{0, h}|_{\boldsymbol{s}_h > h}) \leq \frac{1}{4k \cdot \iter}+\negl(\secp)$.
        \end{proposition}
        \begin{proof}[Proof of Proposition \ref{flooding_three}]
           When $h < \iter$, as noted earlier, it suffices to consider the state at the start of the $h$th iteration of step 5. Observing that this state has no dependence on $i$ or $q$ (the external verifier's query), this implies that the sampling of $\Pi$ in both hybrids is drawn from the same uniform distribution over $\calP$. Therefore, we can apply Lemma \ref{lem:new-csmz} and note that by a coupling argument, we have that 
               \[ \TD(\calH_{0, h-1}|_{\boldsymbol{s}_{h-1} > h}, \calH_{0, h}|_{\boldsymbol{s}_h > h}) \leq \TD(\calD_{h-1}|_{\boldsymbol{s}_{h-1} > h}, \calD_{h}|_{\boldsymbol{s}_h > h}) \leq \frac{\TD(\calD_{h-1}, \calD_{h})}{\Pr[\boldsymbol{s}_{h-1} > h]} \leq \frac{N_{\eps, \delta} \cdot \eta}{\Pr[\boldsymbol{s}_{h-1} > h]}, \]
           and so 
               \[ \Pr[\boldsymbol{s}_{h-1} > h] \cdot \TD(\calH_{0, h-1}|_{\boldsymbol{s}_{h-1} > h}, \calH_{0, h}|_{\boldsymbol{s}_h > h}) \leq N_{\eps, \delta} \cdot \eta \leq \frac{1}{4k \cdot \iter}. \]
           On the other hand, through an identical argument as in Lemma \ref{three_abort}, we observe that
               \[ \Pr[\boldsymbol{s}_{\iter-1} > \iter] = \negl(\secp), \]
           and thus
               \[ \Pr[\boldsymbol{s}_{\iter-1} > \iter] \cdot \TD(\calH_{0, \iter-1}|_{\boldsymbol{s}_{\iter-1} > \iter}, \calH_{0, \iter}|_{\boldsymbol{s}_\iter > \iter}) \leq \Pr[\boldsymbol{s}_{\iter-1} > \iter] \cdot 1 = \negl(\secp), \]
           which finishes the proof.
        \end{proof}
        \begin{proposition}\label{success_three}
            $\TD(\calH_{j, h-1}|_{\boldsymbol{s}_{h-1} = h}, \calH_{j, h}|_{\boldsymbol{s}_{h} = h}) \leq \sqrt{\frac{-\log \xi}{k}}$.
        \end{proposition}
        \begin{proof}[Proof of Proposition \ref{success_three}]
            First, it is easy to see that $\calH_{0, s'-1}$ and $\calH_{0, s'}$ behave identically before the $s'$th iteration. Thus, by a coupling argument, we can fix the state (which does not yet have any dependence on $i$) at the start of the $s'$th iteration.
            Note that since the transcript is fully measured upon success and both hybrids use a query $\bar{q}$ drawn from the same distribution (given a quantum state which has no dependence on $i$) and first round message $\bar{z}_1$, it suffices to bound the trace distance of $(i, \tau)$ and $(j, \tau)$ conditioned on success in the $s'$th iteration, where $\tau$ is the transcript outputted by hybrid $\calH_{0, s'}$.\footnote{We could equally have considered the transcript outputted by hybrid $\calH_{0, s'-1}$, but note that they would behave identically in the $s'$th iteration.}

            To this end, using another coupling argument, it suffices to consider for two fixed indices $i, j \gets [k]$, fixed transcript $\tau$, and fixed internal randomness $\bar{r}$, the potential difference in verdicts of $\SoftDecision_{\nu, t}(i, \bar{r}^{(-i)}, \tau, \omega)$ and $\SoftDecision_{\nu, t}(j, \bar{r}^{(-j)}, \tau, \omega)$. In the case where $i = j$, they behave identically, so we assume that $i \neq j$. Then, note that if $\sum_{n \neq i, j} \Accept_n(\bar{r}, \tau) = \ell$, we have that 
            \[
                \ell+1 \leq 1+\sum_{n \neq i} \Accept_n(\bar{r}, \tau), 1+\sum_{n \neq j} \Accept_n(\bar{r}, \tau) \leq \ell+2,
            \]
            and so for random $i, j \gets [k]$
            \[
                \left|\Pr_{\omega}[\SoftDecision_{\nu, t}(i, \bar{r}^{(-i)}, \tau, \omega) = 1] - \Pr_{\omega}[\SoftDecision_{\nu, t}(j, \bar{r}^{(-j)}, \tau, \omega) = 1]\right| \leq 1-2^{-\nu} \leq \nu.
            \]
            Since the probability of overall acceptance (i.e. the $\SoftDecision$ projection returning 1) is the same for random $i$ as it is for random $j$, we conclude that the trace distance of the outputs of the two hybrids conditioned on success is at most $\nu = \sqrt{\frac{-\log \xi}{k}}$.
        \end{proof}
        Thus, we have that 
        \begin{align*}
            \TD(\calH_{0, h-1}, \calH_{0, h}) &\leq \Pr[\boldsymbol{s}_{h-1} = h] \cdot \TD(\calH_{0, h-1}|_{\boldsymbol{s}_{h-1} = h}, \calH_{0, h}|_{\boldsymbol{s}_{h} = h}) \\
            &+ \Pr\left[\boldsymbol{s}_{h-1} > h\right] \cdot \TD(\calH_{0, h-1}|_{\boldsymbol{s}_{h-1} > h}, \calH_{0, h}|_{\boldsymbol{s}_{h} > h}) \\
            &\leq \Pr[\boldsymbol{s}_{h-1} = h] \cdot \sqrt{\frac{-\log \xi}{k}} + \frac{1}{4k \cdot \iter} + \negl(\secp),
        \end{align*}
        as claimed, concluding the proof of Proposition \ref{three_TD_hybrid}.
    \end{proof}
This concludes the proof of Lemma \ref{hybrids_three_sd}.
\end{proof}

For complete transcripts $\tau$, we define $\bar{r}_{\tau}$ to be the internal randomness of the $k$-fold verifier which is consistent with $\tau$. Similarly, $\bar{r}^{(-j)}_{\tau}$ refers to the randomness of all but the $j$th coordinate of the $k$-fold verifier which is consistent with $\tau$.
\begin{lemma}\label{three_abort}
    $\underset{(i, \tau) \gets \calH_1, \omega}{\Pr}[\exists j: \SoftDecision_{\nu, t}(j, \bar{r}^{(-j)}_{\tau}, \tau, \omega) = 1] \geq 1-\underset{(i, \tau) \gets \calH_1}{\Pr}[\tau = \bot]$.
\end{lemma}
\begin{proof}[Proof of Lemma \ref{three_abort}]
    Like in the proof of Lemma \ref{public_abort}, it suffices to prove that if $\calH_1$ does \emph{not} abort and returns a transcript $\tau$, then for the $j$ and $\bar{r}$ corresponding to the last projection computed in step 5(d) (that is, $\bar{r}$ is the completion of $\bar{r}^{(-j)}$ with the randomness used to generate the associated query $q$), $\SoftDecision_{\nu, t}(j, \bar{r}^{(-j)}, \tau, \omega) = 1$ and $\bar{r}$ is by definition the verifier randomness which is consistent with $\tau$. This follows directly from the observation that step 6 of $\calH_1$ undoes step 5 of the last (successful) $\SoftDecisionProj$ projection and so we can apply Fact \ref{fact:deferring_measurement}.
\end{proof}

Finally, define $\mathsf{Good}_{\nu, t, j} := \{(\tau, \omega) \mid \SoftDecision_{\nu, t}(j, \bar{r}^{(-j)}_{\tau}, \tau, \omega) = 1 \}$ to be the set of transcripts $\tau$ and randomness $\omega$ such that the $\SoftDecision$ function returns 1 on $\tau$ given randomness $\omega$. Then we have the following:
\begin{lemma}\label{soft_decision}
    For all $\nu > 0$, $t = t(\secp)$, and $j$, 
        \[ \Pr_{(i, \tau) \gets \calH_1, \omega}\big[\Accept_i(\bar{r}_{\tau}, \tau) = 1 \mid (\tau, \omega) \in \mathsf{Good}_{\nu, t, j})\big] \geq \frac{t}{k}-\frac{\log k}{\sqrt{k}}-2\sqrt{\frac{-\log\xi}{k}}-\frac{2}{\sqrt{k}}-\negl(\secp). \]
\end{lemma}
\begin{proof}[Proof of Lemma \ref{soft_decision}]
    Note that we can assume that $k \geq 16$ as otherwise the inequality is trivially satisfied. Consider the modified function $\SoftDecision'_{\nu, t}$ which on input $\bar{r}_{\tau}$, $\tau$, and $\omega$, returns 1 with probability $\min\{1, 2^{-\nu(\ell-t)}\}$ where $\ell = \sum_i \Accept_i(\bar{r}, \tau)$. We first consider the difference in the probability that $\SoftDecision'_{\nu, t}$ returns 1 and the probability that $\SoftDecision_{\nu, t}(j, \cdot, \cdot, \cdot)$ returns 1 on input $\tau$. It is clear from a coupling argument that for any transcript, the largest possible difference between these two probabilities is $1-2^{-\nu} \leq \nu$, so we have that
        \[ \Pr_{(i, \tau) \gets \calH_1, \omega}[\SoftDecision'_{\nu, t}(\bar{r}_{\tau}, \tau, \omega) = 1 \mid (\tau, \omega) \in \mathsf{Good}_{\nu, t, j}] \geq 1-\nu. \]
    Thus, it suffices to bound
      \[ \Pr_{(i, \tau) \gets \calH_1, \omega}\big[\Accept_i(\bar{r}_{\tau}, \tau) = 1 \mid \SoftDecision'_{\nu, t}(\bar{r}_{\tau}, \tau, \omega) = 1 \big], \]
    since $\Pr_{(i, \tau) \gets \calH_1, \omega}\big[\Accept_i(\bar{r}_{\tau}, \tau) = 1 \mid (\tau, \omega) \in \mathsf{Good}_{\nu, t, j}\big]$ is at least the product of these two lower bounds.
    
    Here, we use the following lemma from \cite{HPPW10}:
    \begin{lemma}[\cite{HPPW10}]
    Let $D_1, \ldots, D_k$ be binary random variables where $\Pr[\sum_{i=1}^k D_i \geq t] \geq \eps$, let $L = \sum_{i=1}^k D_i$, let $t \leq k$, $\nu > 0$, and let $W$ be a binary random variable such that $\Pr[W = 1 \mid L = \ell] = \min(1, 2^{\nu(\ell-t)})$. Then
    \[
        \frac{1}{k}\sum_{i=1}^k \Pr[D_i = 0 \mid W = 1] \leq 1-\frac{t}{k}+\frac{1}{k\nu}(\log k-\log\eps)+\frac{4}{\nu^2k^2}.
    \]
    \end{lemma}
    \noindent
    Setting $\nu = \sqrt{-\log \xi/k}$ like in our reduction and letting $D_i := \Accept_i$, we first recall that in every iteration of step 5 of $\calH_1$, the $\SoftDecisionProj$ projection in step 5(d) always has a minimum success probability of $2\eps_0 = \xi/2$ for sufficiently large $\secp$, we find that 
    \[
        \frac{1}{k}\sum_{i=1}^k \Pr_{(i, \tau) \gets \calH_1, \omega}[\Accept_i(\bar{r}_{\tau}, \tau) = 0 \mid \SoftDecision'_{\nu, t}(\bar{r}_{\tau}, \tau, \omega) = 1] \leq 1-\frac{t}{k}+\frac{\log k}{\sqrt{k}}+\sqrt{\frac{-\log(\xi/2)}{k}}+\frac{4}{k}+\negl(\secp)
    \]
    Since $\tau$ is independent of $i$ in $\calH_1$, this means that 
    \begin{align*}
        \Pr_{(i, \tau) \gets \calH_1, \omega}[\Accept_i(\bar{r}_{\tau}, \tau) = 1 \mid \SoftDecision'_{\nu, t}(\bar{r}_{\tau}, \tau, \omega) = 1] &\geq \frac{t}{k}-\frac{\log k}{\sqrt{k}}-\sqrt{\frac{-\log\xi}{k}}-\frac{1}{\sqrt{k}}-\frac{4}{k}-\negl(\secp) \\
        &\geq \frac{t}{k}-\frac{\log k}{\sqrt{k}}-\sqrt{\frac{-\log\xi}{k}}-\frac{2}{\sqrt{k}}-\negl(\secp).
    \end{align*}
    Multiplying these two lower bounds, we conclude that 
    \begin{align*}
        \Pr_{(i, \tau) \gets \calH_1, \omega}\big[\Accept_i(\bar{r}_{\tau}, \tau) = 1 \mid (\tau, \omega) \in \mathsf{Good}_{\nu, t, j}\big] &\geq (1-\nu)\left(\frac{t}{k}-\frac{\log k}{\sqrt{k}}-\sqrt{\frac{-\log\xi}{k}}-\frac{2}{\sqrt{k}}-\negl(\secp)\right) \\
        &\geq \frac{t}{k}-\frac{\log k}{\sqrt{k}}-\sqrt{\frac{-\log\xi}{k}}-\frac{2}{\sqrt{k}}-\nu-\negl(\secp) \\
        &= \frac{t}{k}-\frac{\log k}{\sqrt{k}}-2\sqrt{\frac{-\log\xi}{k}}-\frac{2}{\sqrt{k}}-\negl(\secp),
    \end{align*}
    as desired.
\end{proof}
Consequently, from Lemmas \ref{final_three_success}, \ref{hybrids_three_sd}, \ref{three_abort}, and \ref{soft_decision}, it follows that
\begin{align*}
    \Pr[\langle(\calA, \widetilde{V})(\secp, x)\rangle = 1] &= \Pr_{(i, \tau) \gets \calH_0}[\Accept_i(\bar{r}_{\tau}, \tau) = 1] \\
    &\geq \Pr_{(i, \tau) \gets \calH_1}[\Accept_i(\bar{r}_{\tau}, \tau) = 1] - \TD(\calH_1, \calH_0)\\
    &\geq \left(\frac{t}{k}-\frac{\log k}{\sqrt{k}}-2\sqrt{\frac{-\log\xi}{k}}-\frac{2}{\sqrt{k}}-\negl(\secp)\right) \left(1-\Pr_{(i, \tau) \gets \calH_1}[\tau = \bot]\right) - \TD(\calH_1, \calH_0) \\
    &\geq \left(\frac{t}{k}-\frac{\log k}{\sqrt{k}}-2\sqrt{\frac{-\log\xi}{k}}-\frac{2}{\sqrt{k}}-\Pr_{(i, \tau) \gets \calH_1}[\tau = \bot]-\negl(\secp)\right) - \TD(\calH_1, \calH_0) \\
    &\geq \left(\frac{t}{k} - \frac{\log k}{\sqrt{k}} - 2\sqrt{\frac{-\log\xi}{k}} - \frac{2}{\sqrt{k}} - \frac{1}{2k} - \negl(\secp)\right) - \frac{1}{4k} - \sqrt{\frac{-\log\xi}{k}} - \negl(\secp) \\
    &\geq \frac{t}{k}-\frac{\log k}{\sqrt{k}}-\frac{3}{\sqrt{k}}-3\sqrt{\frac{-\log\xi}{k}}-\negl(\secp).
\end{align*}
Since we can assume without loss of generality that $k > 8$ (as $\calA$ always suceeds with non-negative probability), we have that
\begin{align*}
    \Pr[\langle(\calA, \widetilde{V})(\secp, x)\rangle = 1] &\geq \frac{t}{k}-\frac{\log k}{\sqrt{k}}-\frac{3}{\sqrt{k}}-3\sqrt{\frac{-\log\xi}{k}}-\negl(\secp) \\
    &\geq \frac{t}{k}-\frac{2\log k}{\sqrt{k}}-3\sqrt{\frac{-\log\xi}{k}}-\negl(\secp),
\end{align*}
concluding the proof of Theorem \ref{three_message_reduction}.
\end{proof}

\section{Acknowledgments}\label{sec:acknowledgments}
We would like to deeply thank Rachel Zhang and Fermi Ma for their many valuable discussions and perspectives, as well as useful comments and suggestions which simplified the presentation of our results. We would also like to thank Alex Lombardi for his insights. 

This material is based upon work supported by the Defense Advanced Research Projects Agency (DARPA) under Contract No. HR0011-25-C-0300. Any opinions, findings and conclusions or recommendations expressed in this material are those of the author(s) and do not necessarily reflect the views of the Defense Advanced Research Projects Agency (DARPA).

\newpage
\bibliographystyle{alpha}
\bibliography{references.bib}

\end{document}